%% file: arXiv.DRAFT.September.2020.tex
\documentclass[conference,compsoc]{IEEEtran}
\usepackage{amsmath,amssymb,amsfonts}
\usepackage{algorithmic}
\usepackage{lipsum}
\usepackage{graphicx}
\usepackage{caption}
\usepackage{xcolor}
\usepackage{subcaption}
\usepackage[lined,boxed]{algorithm2e}
\usepackage{enumitem,kantlipsum}
\DeclareCaptionFont{xipt}{\fontsize{8}{10}\bfseries}
\usepackage[font=xipt,labelfont=bf]{caption}
\usepackage[justification=centering]{caption}
\usepackage{textcomp}
\SetAlCapSkip{1em}
\usepackage{disclaimer.arXiv}
\usepackage{draftwatermark}
\SetWatermarkText{Draft}
\SetWatermarkScale{1}

\ifCLASSINFOpdf
\else
\fi
\begin{document}
\title{Coverage and Energy Analysis of Mobile Sensor Nodes in\\ Obstructed Noisy Indoor Environment: A Voronoi-Approach}

\author{
\IEEEauthorblockN{K. Eledlebi\IEEEauthorrefmark{1}, D. Ruta\IEEEauthorrefmark{2}, H. Hildmann\IEEEauthorrefmark{3}\IEEEauthorrefmark{2}, F. Saffre\IEEEauthorrefmark{5}\IEEEauthorrefmark{2}, Y. Al Hammadi\IEEEauthorrefmark{1}, and A.F. Isakovic\IEEEauthorrefmark{1}\IEEEauthorrefmark{4}\IEEEauthorrefmark{8}}
%
\IEEEauthorblockA{\IEEEauthorrefmark{1} Khalifa University / \IEEEauthorrefmark{2}Emirates ICT Innovation Centre (EBTIC),
Abu Dhabi, P.O. 127788, UAE;\\email: \{khouloud.edlebi / dymitr.ruta / yousof.alhammadi\}@ku.ac.ae}
\IEEEauthorblockA{\IEEEauthorrefmark{5} Technical Research Centre of Finland (VTT), Vuorimiehentie 3, 02044 Espoo, FI; email: fabrice.saffre@vtt.fi}
\IEEEauthorblockA{\IEEEauthorrefmark{3} Nederlandse Organisatie voor Toegepast-Natuurwetenschappelijk Onderzoek (TNO), \\Oude Waalsdorperweg 63, 2597 AK, The Hague, NL; hanno.hildmann@tno.nl}
\IEEEauthorblockA{\IEEEauthorrefmark{8} Colgate University, Ho Science Center, Oak Dr. 13, 13346 Hamilton, NY; email: aisakovic@colgate.edu}
\IEEEauthorblockA{\IEEEauthorrefmark{4} corresponding author:  A.F. Isakovic -- email: iregx137@gmail.com}
}
\maketitle
\thispagestyle{plain}
\pagestyle{plain}
\begin{abstract}
The rapid deployment of wireless sensor network (WSN) poses the challenge of finding optimal locations for the network nodes, especially so in (i) unknown and (ii) obstacle-rich environments. This paper addresses this challenge with BISON (Bio-Inspired Self-Organizing Network), a variant of the Voronoi algorithm. In line with the scenario challenges, BISON nodes are restricted to (i) locally sensed as well as (ii) noisy information on the basis of which they move, avoid obstacles and connect with neighboring nodes. Performance is measured as (i) the percentage of area covered, (ii) the total distance traveled by the nodes, (iii) the cumulative energy consumption and (iv) the uniformity of nodes’ distribution. Obstacle constellations and noise levels are studied systematically and a collision-free recovery strategy for failing nodes is proposed. Results obtained from extensive simulations show the algorithm outperforming previously reported approaches in both, convergence speed, as well as deployment cost.
\end{abstract}

\IEEEpeerreviewmaketitle
\section{Introduction}
Wireless Sensor Networks (WSN) are used pervasively in the context of e.g., the monitoring of  physiological and environmental parameters, for object tracking and surveillance or in smart homes \cite{4,6,9}. WSNs are commonly optimized with regard to connectivity, coverage, localization, lifetime and dynamic adaptation \cite{4,1,2,3}.

Since the surrounding environment affects the localization of sensor nodes and impedes the data transfer between nodes \cite{1}, the robustness of WSN against noise-driven disruptions is a relevant performance measure. Moreover, WSNs are often supported by low power devices, making energy efficiency a factor that possibly affects e.g., the lifetime of the network or the complexity of the algorithms.

The efficient deployment of sensor nodes (fast and cheap while delivering maximum service) is a main issue \cite{10,11,14}, especially in scenarios where this has to happen with limited or no prior knowledge about the target environment.

To facilitate rapid deployment, a common approach is to move nodes into an environment continuously (on-the-fly) and based on what they can sense in real-time \cite{13,15}.

\begin{figure}[h]
\centerline{\includegraphics[width=0.8\hsize]{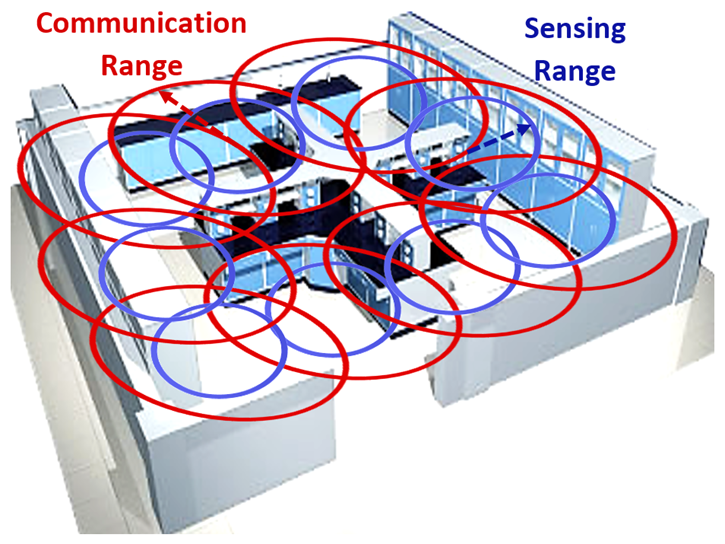}}
\caption{An example of a WSN in a small laboratory. Each sensor node has two ranges: the smaller, sensing range $R_S$ (shown as blue circles), and the larger, communication range $R_C$ (the red circles).}
\label{intro}
\end{figure}

In this paper,
    we propose a Voronoi-based algorithm called BISON,
    (for Bio-Inspired Self-Organizing Network),
        for providing nearly optimum solution to
            self-deployment and
            localization
        of a WSN in
            an unknown bounded region of interest
                with obstructions such as in Fig\ref{intro},
            in the presence of noise distortion.

BISON evolves the positioning of the sensor nodes inside
an area, making use of their distribution to maximize the coverage and maintain the connectivity through controllable injection conditions. To facilitate energy conservation we compute the nodes' next positions iteratively and based solely on local information (i.e., about a node's neighbours within its sensing range) \cite{16,17}. BISON's performance improves in the presence of various levels of Gaussian noise, which makes it useful for implementations.

In \S \ref{sec:overview.voronoi} we briefly discuss existing Voronoi-based algorithms for WSN and highlight the novelty in our self-organizing network (BISON) (introduced in \S \ref{sec:BISON}). Our model and its implementation is provided in \S \ref{sec:modelling}. In \S \ref{sec:evaluation} we define our evaluation metrics for comparison of the proposed approach with representative competitive algorithms, the parametric performance analysis is discussed in \S \ref{sec:results}.
\section{Voronoi-based Algorithms for WSN}\label{sec:overview.voronoi}
\subsection{Inspiration and state of the art}
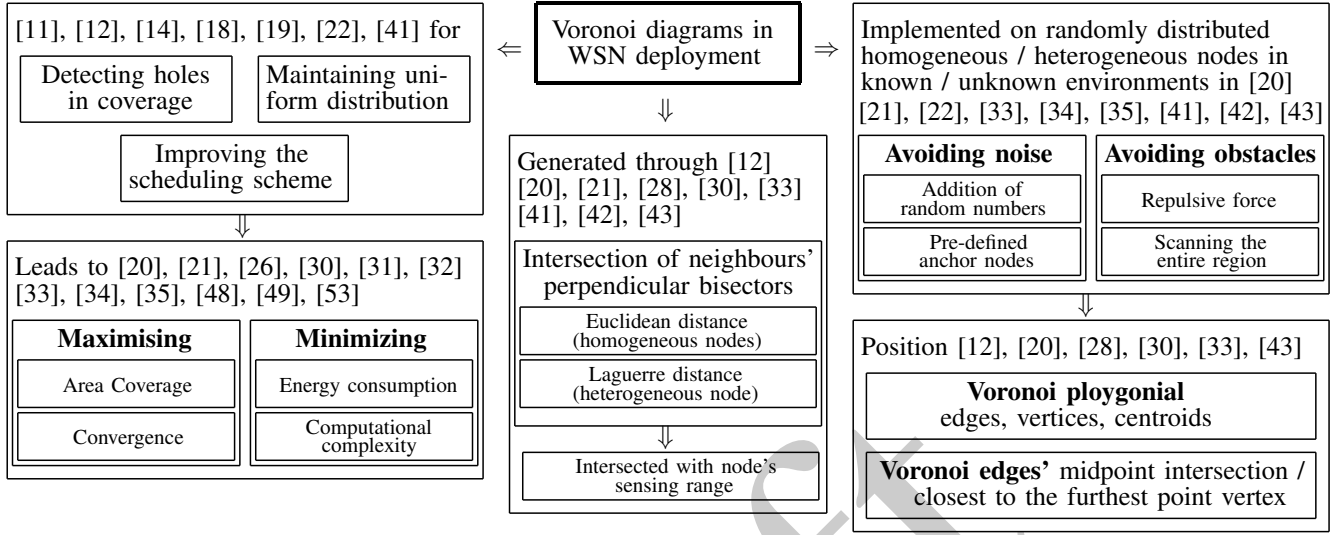
\begin{figure*}[h]
\input{diagram}
\caption{A schematic representing an overview of Voronoi-based approaches goals and techniques used during wireless sensor network deployment.}
\label{flowchart}
\end{figure*}

Animal strategies of self-organization 
to reach a dynamic steady state distribution have inspired the application of such behaviors in mobile networks \cite{15,21,22,23}. A known benefit is the ability to operate with a minimal amount of information about the environment, yet to achieve near optimal solutions with low(er) computational demands. Whether they are considered a geometrical phenomenon \cite{24}, or a bio-inspired one \cite{16,36,37}, Voronoi tessellations are among such techniques.
Fig. \ref{flowchart} offers a brief schematic review for Voronoi-based algorithms goals and techniques, for more details we refer to Fig. S1  of the Supporting Online Materials (SOM, hereafter).
%
Territorial behaviour in animals such as for the male \textit{tilapia mossambica} (a fish species) and how polygonal shapes of sufficient density are formed is discussed in  \cite{37}; an illustration is found in Fig. S2 in SOM.

By mimicking such behavior, sensor nodes in a WSN can generate Voronoi regions (convex polygons that together partition an area) based on local information and subsequently adjust their positions until the change per iteration falls below a system convergence condition. The boundaries of the polygon are generated by drawing the perpendicular bisector of each segment connecting the cell node with its neighbor nodes (for a pictorial representation of this cf. Fig. \ref{noise} on page \pageref{noise} or Fig. S2 in SOM).

This process tends to achieve uniform distribution during the self-organizing process, motivating WSN researchers to analyze and identify the coverage holes after the initial distribution of sensors. Voronoi algorithms have been used to enhance network-coverage and -connectivity, either stand-alone \cite{40,41,42} or in tandem with bio-inspired algorithms \cite{28,29,32,48,51}.

For known areas and a fixed number of homogeneous sensor nodes, Wang \textit{et al.} \cite{38,IEEE-j1,39} developed three different coverage-optimizing algorithms which re-allocate the nodes to positions covering most of the Voronoi vertices/edges within the sensors' sensing ranges. Apart from the vertices, there is a substantial interest towards approaching the centroid (center of mass of a Voronoi cell) during sensors' re-allocation, known to provide better uniform node-distribution while requiring less movement \cite{40,41,42,42a}. An example with the use of a Voronoi centroid is discussed by Zou \textit{et al.} \cite{43,44,45,46} who introduced Node Spreading Voronoi Algorithm (NSVA) to self-deploy fixed number of nodes in an unknown obstacle-free area.

Voronoi diagrams are also involved in WSN deployment of heterogeneous sensors, whether combined with Laguerre geometry to partition the regions between adjacent sensors, or adding weights on each sensor node based on its sensing range to create multiplicatively-weighted Voronoi diagrams \cite{46a,46b,46c,46cc,46d}.  Additionally, Voronoi diagrams are involved in the scheduling mechanisms of WSN \cite{42}, and are considered in security algorithms, providing robustness against malicious attacks that might affect the nodes identities during the deployment process \cite{46f}.
\subsection{Novelty of BISON}
Previous discussions of Voronoi-based algorithms considered a random distribution of sensor nodes, with most algorithms relying on prior knowledge about the (often obstacle-free) environment and node locations while assuming a guaranteed connectivity among the sensor nodes.
Relying on first scanning the entire environment to determine obstacle location before distributing nodes is putting a strain network resources such as time and energy \cite{46ba, 46bb}.

In contrast, our approach evolves the location of homogeneous sensor nodes in  unfamiliar and obstructed noisy environments. This happens iteratively, allowing the nodes to investigate their injection process to direct themselves based on methodological conditions to achieve maximum coverage and to maintain connectivity in the network, without a priory assumption. In BISON, each nodes scans only its local environment to identify nearby obstacles, generates or updates its own Voronoi region (accounting for obstacles) and then directs itself toward its new Voronoi centroid.

In addition, the negative impact of communication noise on the distribution of nodes is addressed in the literature either by using anchor nodes of known locations to predict the location of all other nodes (using fuzzy logic methods), or by adding a (random) uncertainty value to the measured distances between the nodes and their neighbors. In our work we simulated realistic communication noise (Gaussian) and embedded this in the information about nodes' locations (used extensively in the deployment process).

Finally, we will show (a) that involving Voronoi-based centroidal WSN node deployment in BISON is efficient in the removal of dead / tired nodes with least distance traveled among the nodes, as well as (b) how underlying kinematics of WSN nodes are affected by environmental noise that lead to improvements in the network coverage. The latter we demonstrate in our noise coherence analysis.
\subsection{Contribution of this article}
Our work is guided by three main goals:
\begin{enumerate}
    \item \textit{Propose an approach that can handle the inclusion of obstacles and is capable of examining their effect.}

    \item \textit{Examine / evaluate the performance of the proposed approach when subjected to noise.}

    \item \textit{Ensure resilience against uncontrollable loss of WSN nodes and facilitate controlled node removal / exiting of low energy nodes by providing individual exit paths.}
\end{enumerate}
~\\
The main contributions to, and improvements \textit{over}, the state of the art are that
we evolve sensor nodes' locations without
(1a) \textit{a priori} fixing their number or
(1b) relying on special injection conditions to ensure network coverage and connectivity; that the approach is designed for unknown environments
(2a) with obstacles but
(2b) without a prior demand on a continuous connected network;
(3) that the system is subjected to communication noise. We also
(4) show that the swarm coherence method is an effective tool.

\section{The BISON Algorithm}\label{sec:BISON}
The proposed algorithm evolves network nodes inside an environment, following certain methodological conditions that aim for optimum coverage while maintaining connectivity without \textit{a priory} assumption. A centroid Voronoi tessellation mechanism is implemented for the deployment of nodes in an unknown noisy environment with obstacles. The number of nodes involved in the deployment process is unknown (but finite) and dynamically determined by gradual and conflict-free absorption. The node injection conditions are designed to ensure full coverage and connectivity. Node movement is achieved through self-organization, with the new allocation of nodes guided by the nodes' desire to move towards the respective center of their Voronoi cell; cf. Fig. S3 in SOM for a flowchart of the BISON algorithm.
\subsection{Voronoi regions in obstacle free environments}\label{subsec:BISON.plain}

Unlike in prior art, the Voronoi regions in BISON are are not only characterized by the neighboring nodes, which might drop the connection in noisy environment or drop out altogether unexpectedly.
Instead, the configuration is dependent on observations within a sensor's sensing range; the Voronoi regions are
the intersection between the neighboring bisectors, the obstacles, and the sensor's sensing range.

For starters, any point \ensuremath{x} inside the environment \ensuremath{A} is assigned to the region \ensuremath{V^{*}_i} of a specific node \ensuremath{n_i} (cf. \cite{36}):
\begin{equation} \label{eq:2}
V^{*}_i = {\{x| \forall n_j \in N\backslash \{n_i\}: |d(x,n_i)| < |d(x,n_j)| \}}
\end{equation}
Practically, the allocations defined by Equation \ref{eq:2} are formed continuously through local interaction: the nodes perpendicular bisectors are calculated using the Euclidean distance \ensuremath{d(from,to)} to neighbouring (i.e., within communication range) nodes. The more restricted BISON Voronoi regions \ensuremath{V_i} are determined using the intersection between the perpendicular bisectors and the node's sensing range (\ensuremath{V_i\subseteq \ensuremath{V^{*}_i}}).

For nodes at the outer rim of the network (in Fig.\ref{bison1} (c): e.g., $n_1$, $n_2$ and $n_3$) their Voronoi region \textit{is} their sensing range: \ensuremath{V_{i} = R_{i S}}. These nodes are temporarily static (as they already are at their Voronoi centroid) and disconnected from the network. As additional nodes are injected, these outer nodes will be reached and pushed out further.
\subsection{Node injection}\label{subsec:BISON.node.injection}
With regard to additional nodes being added, this is conditional on either of the following two conditions:

\begin{enumerate}
\item there are disconnected nodes, or\hfill(\texttt{c1})
\item there is unexplored space left.\hfill(\texttt{c2})\label{crossreference:injection.conditions}
\end{enumerate}

To formalize this, we first impose the requirement of connectivity, i.e., that every node has at least one neighbouring node within its communication range $R_C$:
\begin{equation*}\label{eq:3}
\ensuremath{ \forall n_i \in N, \exists \  n_j \in N\backslash\{i\} : |d_t(n_i,n_j)| \leq \sqrt{3}R_S}
\end{equation*}
If anode is violating this then condition (\texttt{c1}) is met and an additional node is injected. Practically, this is implemented by maintaining an account of the number of injected nodes and comparing this to the nodes currently relaying information in the network (a known / knowable property).

Secondly, we require that there is at least one node within sensing range of the point of injection:
\begin{equation*}\label{eq:4}
\ensuremath{ \exists \  n_i \in N : |d_t(n_i,(0,0))| \leq R_S}
\end{equation*}
Assuming that the injection point is at the border of the area, the diffusion of the nodes across the area will cause the node closest to the injection point to drift away if there are uncovered locations anywhere in the area. Therefore, if the above is violated then we can conclude that there are uncovered areas (amongst them the injection point), which violates condition (\texttt{c2}) and causes another node to be injected. Jointly, these two conditions ensure coverage and connectivity; furthermore, in any finite area these conditions can be realized with a finite number of nodes.
\subsection{Node movement}\label{subsec:BISON.node.movement}
Nodes continuously move towards their Voronoi centroids (the center of mass of each node's Voronoi region), thereby ensuring gradual and directed movement towards complete area coverage \cite{58}. This process requires first to find the center of mass of each node's Voronoi region \ensuremath{V}. Any region \ensuremath{V} is a polygon with vertex coordinates \ensuremath{(x_{1},y_{1}), \ldots, (x_{n},y_{n})}, and we calculate its area and center of gravity using measure \ensuremath{m(i) = (x_i \times y_{i+1} - x_{i+1} \times y_i)} (with \ensuremath{x_{n+1} = x_1} and \ensuremath{y_{n+1} = y_1} to come full circle) \cite{cortes}:
\begin{equation*}
Area = \frac{1}{2}\sum_{i=1}^{n}m(i)
\end{equation*}
\textit{Area} is the area of polygon \ensuremath{V}. Once we have this, the x-/y-coordinates of its center of gravity can be found as follows:
\begin{equation*}
C x = \frac{1}{6 \times Area}\sum_{i=1}^{n}(x_i + x_{i+1}) \times m(i)
\end{equation*}
\begin{equation*}
C y = \frac{1}{6 \times Area}\sum_{i=1}^{n}(y_i + y_{i+1}) \times m(i)
\end{equation*}

The distance to its center of gravity, i.e., the difference between the Voronoi centroid (\ensuremath{C_x, C_y}) and the node's actual location (\ensuremath{n_x, n_y}) - together with application specific parameter \ensuremath{\tau} can be used as termination condition. If the aggregated movement of the node over a number of steps does not exceed \ensuremath{\tau} then the algorithm terminates. This is measured over all nodes (cf. Alg. \ref{alg}, below) or, for true decentralization, for each node individually.

\begin{algorithm}[h]
\SetAlgoLined
\caption{Pseudo-code for the BISON algorithm.}
\label{alg}
Initialization\;
define \texttt{min\ensuremath{_{\texttt{PAC}}}} ~~~~~~~~~~~~~~~~~~~~~~~~~~~~~~(\ensuremath{\Leftarrow} cf. \S \ref{subsec:Model.noise.impact})\;
\ensuremath{count \leftarrow 0}; \ensuremath{shift \leftarrow (\tau+1)}\;
\While{(shift $>$ $\tau$ \textnormal{(\S \ref{subsec:BISON.node.movement})}) $||$ count $<$ \texttt{c\ensuremath{_{max}}}) \textnormal{(\S \ref{subsec:Model.noise.impact})}}
    {
    \If{ (injection condition \texttt{c1} or \texttt{c2} \textnormal{(\S \ref{subsec:BISON.node.injection})})}
        {
        inject node at random \ensuremath{\theta} from origin (0,0)\;
        }
    \For{each node \ensuremath{n_i \in N}}
        {
        add random noise ~~~~~~~~~~~~~~~~~(\ensuremath{\Leftarrow} cf. \S \ref{subsec:BISON.noise})\;
        calculate and move towards $C_i$ ~~(\ensuremath{\Leftarrow} cf. \S \ref{subsec:BISON.node.movement})\;
        $\textit{shift}_i$ $\leftarrow$ distance moved ~~~~~~~~(\ensuremath{\Leftarrow} cf. \S \ref{sec:evaluation:DistanceTraveled:implementation})\;
        }
    \ensuremath{count = count +1}; \ensuremath{shift = \sum^N \textit{shift}_i}\;
    calculate \ensuremath{\texttt{PAC}} ~~~~~~~~~~~~~~~~~~~~~~~~(\ensuremath{\Leftarrow} cf. \S \ref{sec:evaluation:PAC:modelling})\;
    \If{ \texttt{PAC} $\geq$ \texttt{min\ensuremath{_{\texttt{PAC}}}} \textnormal{(\S \ref{subsec:Model.noise.impact})})}
        {
        count = 0\;
        \texttt{min\ensuremath{_{\texttt{PAC}}}} $\leftarrow$ \texttt{PAC}\;
        }
    }
\end{algorithm}
\section{Modelling and implementation}\label{sec:modelling}
In this section we provide additional information regarding the model used for our implementation and approach. In addition to the basic modelling choices made we discuss how obstacles and noise were included in the system and how real-world aspects such as node-failure were modelled.
\subsection{Preliminaries and basic modelling choices}\label{subsec:BISON.assumotions}
The following basic modelling choices were made:

\begin{enumerate}
  \item All nodes are homogeneous with equal sensing-, communication-, computation-, and mobility-abilities.
  \item To make the simulations more realistic, the deployment of the sensor nodes happens iteratively, one at a time and from a fixed entry point at the edge (here: \ensuremath{0,0}).
  \item Node-injection is triggered by conditions (cf. \S \ref{subsec:BISON.node.injection}) that verifiable at the entry point; the node's individual entry \textit{angle} (\ensuremath{\theta_i}) is random, between \ensuremath{0^\circ} and \ensuremath{90^\circ}.
  \item Each node has the ability to distinguish its own location and that of other nodes through suitable indicators for indoor environments \cite{58,57,IEEE-j2}.
  \item The communication range (\ensuremath{R_C}) and sensing range (\ensuremath{R_S}) are circular with \ensuremath{R_S = 2} and \ensuremath{R_C = \sqrt{3}R_S}. This relation (cf. \cite{22,56-0,56}) is proven to minimize the overlap between sensing ranges, removing unnecessary (and thus
  inconsequential) complexity.
  \item The threshold for our termination criteria \ensuremath{\tau} was set to one hundredths of the sensing range of a cell: \ensuremath{\tau = \frac{R_S}{100}}
\end{enumerate}
\begin{figure}[h]
\centering
\includegraphics[width=0.55\vsize, height=0.85\hsize]{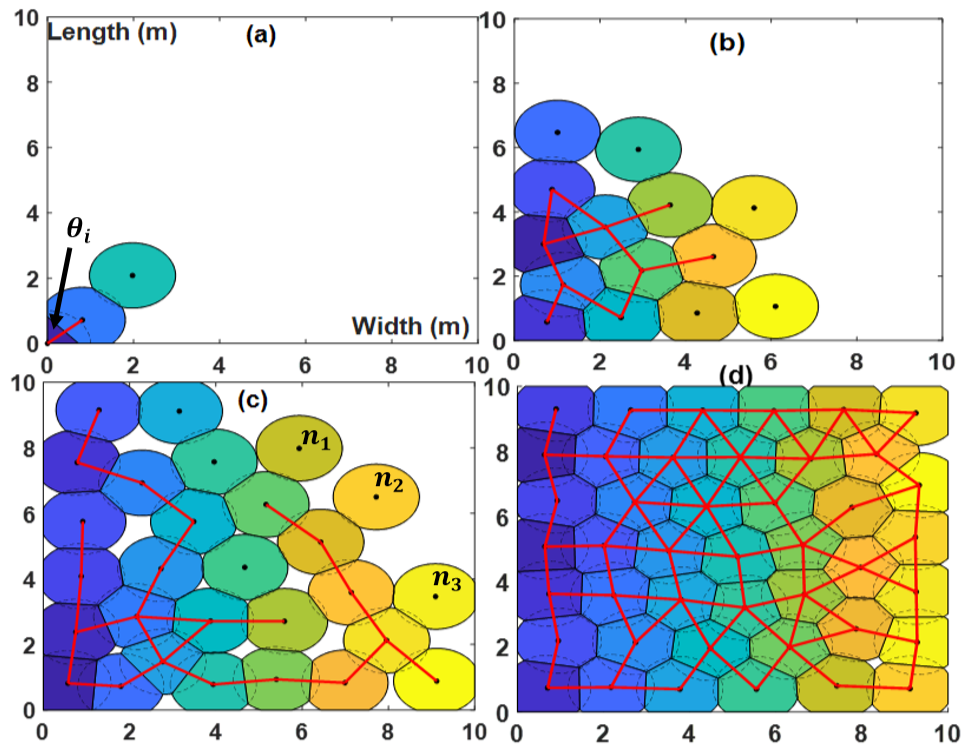}
\caption{Stages of node distribution: nodes iteratively enter the area from the bottom left corner (a) and form Voronoi boundaries (b). This process of covering and connecting continues (c) until the entire area is covered with nodes forming a connected network (d).}
\label{bison1}
\end{figure}
\newpage
\subsection{Introducing obstacles}\label{subsec:BISON.obstracles}
\subsubsection{Motivation}
Virtually all WSNs are deployed in environments with obstacles that obstruct the wireless connection in one way or another. This means that environments such as the one depicted in Fig. \ref{bison1} are over simplified.

\begin{figure}[htbp]
\centerline{\includegraphics[width=\hsize, height=0.95\hsize]{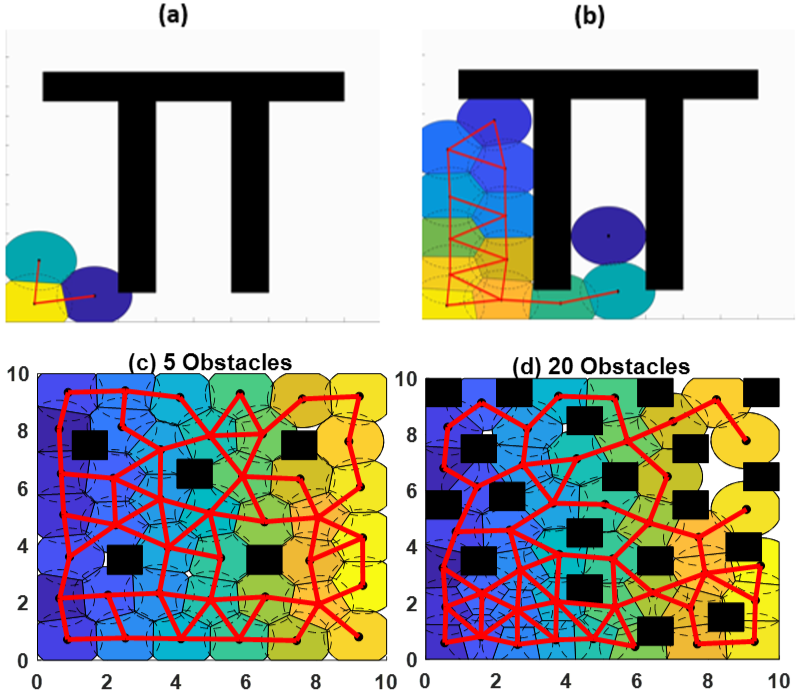}}
\caption{Environments containing walls (top) or pillars (bottom). In addition to blocking signals (which impacts the Voronoi region of a node), walls - and to a lesser effect, pillars -
affect how nodes spread (i.e., move towards the centroid of their Voronoi region).
Environments with pillar-type obstacles contained between 5 (c) and 20 (d) of them while the different wall configurations used are shown in Fig. \ref{shapes}.}
\label{obstacles}
\vskip-2ex
\end{figure}

To account for this we have included signal obstructions (i.e., obstacles) into our model. Figs.
\ref{obstacles}, \ref{shapes} and \ref{spacing} show different aspects of obstacles included in the environment.

\subsubsection{Modelling}
As mentioned in \S \ref{subsec:BISON.assumotions}, we model the sensing range of a node to be circular (omni-directional, transmitting their signal $360^\circ$). Signals bounced back by an obstacle can be used to determine the location of the obstacle \cite{46g}. In our simulation, the calculations required to infer the location of an obstacle is omitted, the relevant information is simply provided when calculating the Voronoi region (which is generated based on the detected obstacle boundaries, the node's sensing range, and the neighboring bisectors).

\begin{figure}[htbp]
\centerline{\includegraphics[width=\hsize, height=0.95\hsize]{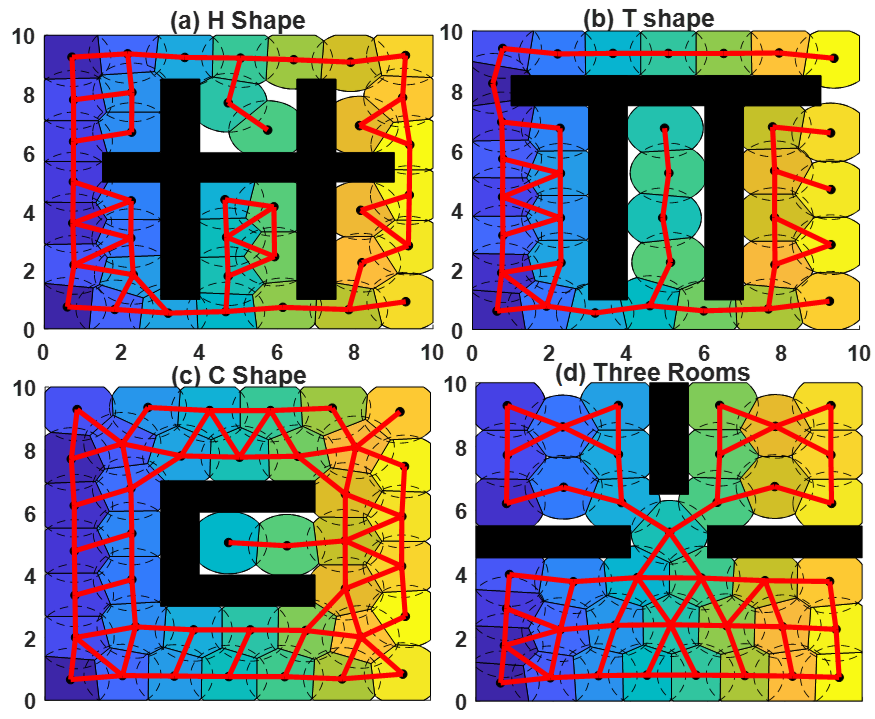}}
\caption{Walled environments. Four variations were used for testing: (a) \textit{``H-shaped''}, (b) \textit{``\ensuremath{\pi}-shaped''}, (c) \textit{``C-shaped''} and (d) \textit{``3 rooms''}.\newline
These 4 contain all challenges we identified when looking at floor plans.}
\label{shapes}
\end{figure}

\begin{figure}[htbp]
\includegraphics[width=1\hsize, height=0.95\hsize]{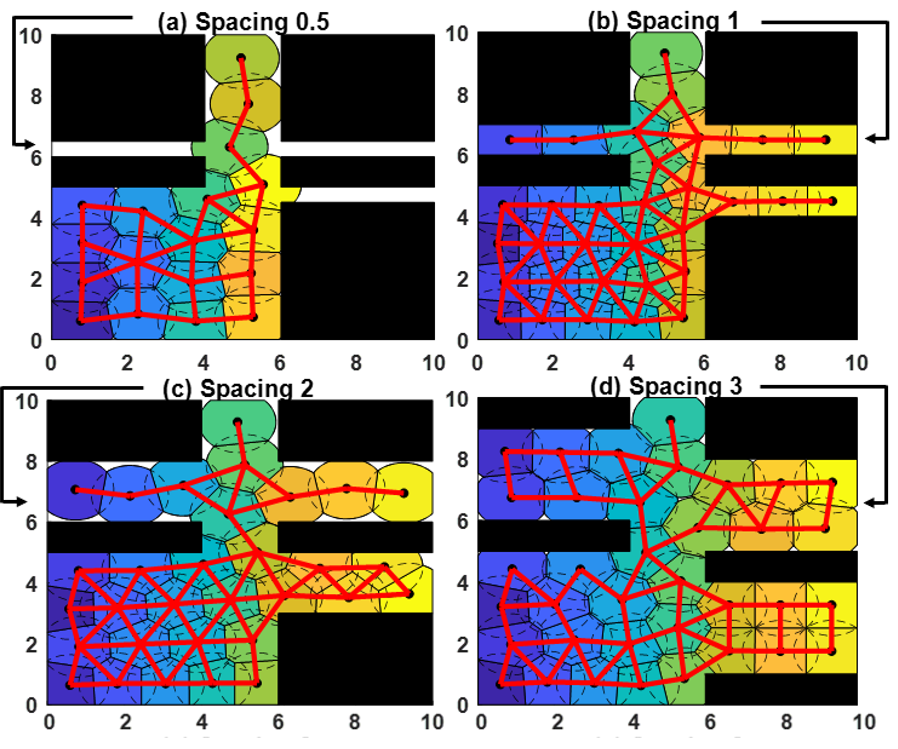}
\caption{Crevices and cavities: we investigated environments where the accessible space is restricted. The width of the crevices ranged from 0.5 (a) to 3 (d) measurement units, which corresponds to roughly 0.25 and 1.5 (respectively) times the sensing range \ensuremath{R_S} of a node.}
\label{spacing}
\end{figure}

We considered three types of obstacles in environments:
\begin{enumerate}
    \item \textbf{small} (1 measurement unit), free-standing obstacles that correspond to pillars or vegetation, cf. Fig. \ref{obstacles} (c) and (d). For the evaluation of the algorithm we simulated environments containing 5 (c) to 20 (d) obstacles.
    \item \textbf{large}, elongated rectangular obstacles representing e.g., walls. 4 different configurations were used, cf. Fig. \ref{shapes}.
    \item elongated \textbf{cavities} as found e.g., between factory equipment, cf. Fig. \ref{spacing}. BISON was tested over a number of variations of this environment, with the distance between obstacles ranging from 0.5 to 3 measurement units (corresponding to \ensuremath{1/4^{th}} to \ensuremath{1/5^{th}} of \ensuremath{R_S}).
\end{enumerate}

The goal was to consider small and large obstacles as well as to explore the ability to deploy WSN into areas with relatively small cavities. The specific instances within each type (cf. Figs. \ref{obstacles}, \ref{shapes} and \ref{spacing}) were, however, relatively freely chosen and are therefore not guaranteed to exhaustively represent all possibly occurring environments.
\subsubsection{Implementation}
The formulae for the calculation of the Voronoi centroid, presented in \S \ref{subsec:BISON.node.movement}, are applicable to all solid polygons. However, while unlikely to happen in our simulation, the calculation for polygons with \textit{``holes''} (pillars) requires extra steps: we have to first decompose the region into smaller, solid regions. Using the calculation for an \textit{Area} from \S \ref{subsec:BISON.node.movement}, the Voronoi centroid (\ensuremath{C x, C y}) of a non-convex polygon was then calculated using the centroids of its \ensuremath{n} sub-polygons (\ensuremath{C x_1, C y_1}), \ensuremath{\dots}, (\ensuremath{C x_n, C y_n}) together with the respective areas \ensuremath{Area_1, \ldots, Area_n} as follows \cite{58}:
\begin{equation*}
C x = \frac{\sum_{i=1}^{n}C x_{i} Area_i}{\sum_{i=1}^{n} Area_i}
\textnormal{~~and~~}
C y = \frac{\sum_{i=1}^{n}C y_{i} Area_i}{\sum_{i=1}^{n} Area_i}
\end{equation*}
\subsection{Adding noise}\label{subsec:BISON.noise}
\subsubsection{Motivation}
As suggested by the name, nodes in a WSN communicates use \textit{wireless} communication. This is unshielded and naturally incurs distortion and noise from the physical environment. In the context of this article, we only concern ourselves with the impact of noise regarding the determination of the position of neighbouring nodes.

In BISON, nodes regularly check up on their neighbours and attempt to determines their distance to them. The accuracy of this process is affected by the noise in the sensed values and the resulting uncertainty will in turn impact the determination of the bisector lines, which are used to calculate the node's Voronoi region.

To visualize this, Fig. \ref{noise} (left) shows the actual locations of nodes neighbouring \textbf{1}, labelled \textbf{2}, as well as (right) the erroneous perceived positions, labelled \textbf{3}. Using node \ensuremath{x_j} (correct location) / \ensuremath{x^{\prime}_j} (erroneous location) as an example we illustrate the impact noisy sensing can have on the calculation of the bisecting lines and subsequently on the resulting Voronoi region. Depending on the level of noise, the latter can thus differ significantly which can lead to a significant loss of overall coverage (cf. Fig. \ref{obf_n}).

In contrast to e.g., \cite{45}, where randomization was used, we wanted to use a specific noise function. This (a) allows us to investigate the performance of BISON under various degrees of noise distortion (see \S \ref{subsec:BISON.noise.implementation}) as well as (b) facilitates subjecting the system to other, environment specific, noise functions (potential future work).
\subsubsection{Modelling}
In communication channels, additive white Gaussian noise (AWGN) is used to describe the noise affecting the communication channel. Noise is added to the signal with uniform power across the transmission channel.

AWGN is the simplest model of Gaussian distribution with zero mean, modelled as statistical noise with a probability density function \textit{P(x)} (normal distribution) as follows:
\begin{equation}
P(x) = \frac{1}{\sigma \sqrt{2\pi}}e^{\frac{(x-\mu)^2}{2\sigma^2}}
\label{eq noise}
\end{equation}
with \ensuremath{\mu} and \ensuremath{\sigma} the mean and standard deviation of random variable \ensuremath{x}, respectively. Increasing variance \ensuremath{\sigma^{2}} (given by $\sigma^2=\frac{N_o}{2}$ with $N_o$ the \textit{noise power}) has an increased (detrimental) effect on signal accuracy.
%
\subsubsection{Implementation}\label{subsec:BISON.noise.implementation}
The effect of \ensuremath{\sigma} is felt only within \ensuremath{R_C}. I.e., visibility of neighbouring nodes is maintained but their derived locations may be subject to error. Three different noise-levels were considered in our simulations:
\ensuremath{\sigma = 0.01} (low),
\ensuremath{\sigma = 0.05} (medium) and
\ensuremath{\sigma = 0.1} (high).

\begin{figure}[h!]
\centerline{\includegraphics[width=0.49\hsize]{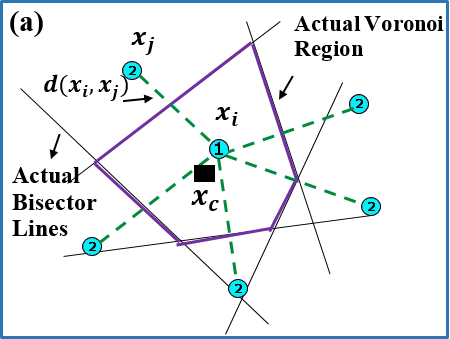}
\includegraphics[width=0.49\hsize]{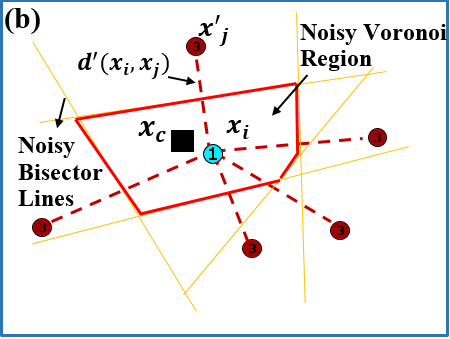}}
\caption{Wireless communication is subject to noise and distortion. This can significantly impact the accuracy of a node's assumption of its neighbours locations. Above, in (a) we depict the actual state of affairs while (b) shows the erroneous locations as well as the resulting Voronoi region. The actual location of nodes is shown in blue / light while erroneously calculated locations are depicted in red / dark.}
\label{noise}
\end{figure}
\subsubsection{Impact}\label{subsec:Model.noise.impact}
\begin{figure}[h!]
\centerline{\includegraphics[width=\hsize, height=0.95\hsize]{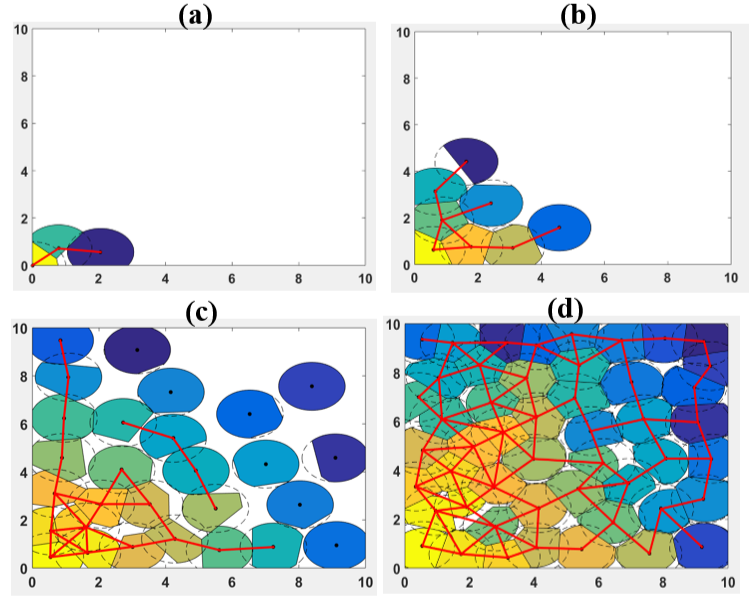}}
\caption{Node-deployment and -movement under noise, compared to Fig. \ref{bison1} without. The effect of calculating incorrect Voronoi centroids due to noise (cf. Fig. \ref{noise}) is clearly visible as the network spreads out. The nodes assume locations that facilitate a lower connectivity between nodes (c) as well as effectively cover a lesser area of the environment (d).}
\label{obf_n}
\end{figure}

Including signal distortion means that incorrect Voronoi centroids are being calculated by the nodes (cf. Fig. \ref{noise}), causing them to move to a position other than the actual center of gravity, as defined by Equation \ref{eq:2}. This in turn can cause incomplete coverage (cf. Fig. \ref{obf_n}, (d)) as well as continued (and unnecessary) consumption of energy as nodes keep moving around their actual Voronoi centroid.

Furthermore, signal distortion can affect the usefulness of the termination condition we use (see next section).
\subsubsection{Additional stop criterion}\label{subsec:Model.noise.2nd.stop}
In noisy environments, and as nodes become increasingly scattered, signal distortions can cause nodes to keep moving back and forth which may result in them exceeding threshold \ensuremath{\tau} (cf. \S \ref{subsec:BISON.node.movement}), used in Alg. \ref{alg} as the first of two termination criteria.
Because of this, a second criteria is devised as follows:
using a pre-defined minimum network coverage area \texttt{min\ensuremath{_{\texttt{PAC}}}} and a constant \texttt{c\ensuremath{_{max}}}  (set in our simulation to \texttt{c\ensuremath{_{max}}}\ensuremath{ = 15}), we stop the algorithm after exceeding \texttt{min\ensuremath{_{\texttt{PAC}}}} and when there was no increase in the coverage area for \texttt{c\ensuremath{_{max}}} consecutive iterations (while the network is fully connected); see Alg. \ref{alg}.
\subsection{Node-failure and self-organizing recovery}\label{subsec:BISON.node.failure}
\subsubsection{Motivation} Having considered real-world challenges in environments (such as obstacles and signal-noise), it would be strange to ignore practical issues of the sensor nodes themselves. Under realistic deployment conditions, a WSN is likely to face sudden / unexpected loss of nodes sooner or later. This may be due to a number of reasons, with the most common examples being device malfunction, resource (battery) depletion or outside influences.
\subsubsection{Modelling}
The modelling of node-failure is straight-forward, in our implementation we can simply remove a node from the WSN. That being said, we would like to distinguish three types of nodes: devices whose Voronoi region touches (1) only other Voronoi regions, (2) one wall or (3) two walls (in which case the node is in a corner).
\subsubsection{Implementation} To enable node-failure handling, no additional implementation is required:
\label{crossreference:remove:003}
losing a node means that some of the remaining nodes lose a neighbour. This means that their Voronoi regions (and thus centroids) change, wich causes them to move. This will happen until coverage is restored. Note that this might cause additional nodes to be injected (but that as well is functionality that is already implemented, cf. \S \ref{subsec:BISON.node.injection}).\label{crossreference:remove:004}

\subsubsection{Additional considerations}
With more complex systems and increasingly autonomous nodes in mind, we briefly touch on the subject of planned / intentional node-loss. This can e.g., be the case when an individual node is low on power and autonomously decides to return to the injection point to recharge. In that case an approach would be to let a node continue to provide coverage until its Voronoi region is entirely taken over by its neighbours. A practical issue would then be the path planning for the now redundant node, which is briefly discussed in \S \ref{crossreference:exit.path.for.redundant.node}.
\section{Evaluation Metrics}\label{sec:evaluation}
In this section, we define the four major metrics used to evaluate the WSN created by BISION: the coverage achieved by the WSN, the combined distance travelled by the nodes to achieve this coverage, the energy consumed to do so and, finally, the uniformity of the node distribution.
\subsection{Percentage Area Coverage (\texttt{PAC})}\label{sec:evaluation:PAC}
\subsubsection{Motivation}
While we are aware of WSN-applications that operate under (sometimes extreme) resource scarcity, our algorithm was designed to achieve full coverage in scenarios of resource abundance, meaning that our concern is \textit{``covering the complete area''} and not \textit{``how to cover as much as possible given a small number of nodes''}. We will therefore use the area covered (in percentage of the environment) as a performance measure and to plot progress during the deployment phase of the WSN.
\subsubsection{Modelling}\label{sec:evaluation:PAC:modelling}
Calculating the \textit{percentage area coverage} (\texttt{PAC}) is done using global insight, i.e., we use the actual coverage of the system  \cite{43,45} and \textit{not} the erroneous assumptions of the individual nodes. Overlapping coverage as well as the impact of walls and pillars are not included:
\begin{equation*} \label{eq:7}
\texttt{PAC} = \frac{\textit{Combined coverage by the WSN}}{\textit{Maximum area that can be covered}}
\end{equation*}
\subsection{Distance Traveled (\texttt{ADT} / \texttt{CDT})}\label{sec:evaluation:DistanceTraveled}
\subsubsection{Motivation}
The energy requirements for a successful deployment of WSN using BISION was an important parameter to consider. While we provide a separate measure for the actual energy consumed (cf. \S \ref{sec:evaluation:EnergyConsumed}) we first want to evaluate performance de-coupled from the energy cost incurred by the operation of the network (i.e., without the cost generated by sensing and signal operations), and we did so by measuring the distance traveled by the nodes.

We measure the distance traveled by the nodes in two ways: we fist consider the simple value of how much, given a current positions of the nodes, each node would have to move if it directly went to the current position in the deployment. This is called the \textit{average distance traveled} (\texttt{ADT}). In addition, we capture the far more accurate value of the \textit{cumulative distance traveled} in the network (\texttt{CDT}), i.e., the 
distance that was actually
traveled, including the noise-induced bouncing back and forth between positions.
\subsubsection{Modelling the average distance traveled (\texttt{ADT})}\label{sec:evaluation:DistanceTraveled:modelling:ADT}
We used the following equation \cite{43,45} to analyze the average distance traveled (\texttt{ADT}) by the sensor nodes:
\begin{equation*} \label{eq:8}
\texttt{ADT} (t) = \frac{1}{n (t)} \sum_{i=1}^{n (t)} |d\big(\textit{origin},(C_i x, C_i y)\big)|
\end{equation*}
where \textit{n(t)} is total number of nodes at time \textit{t}, \textit{origin} is the initial location / the injection point of node \textit{i} (for us that is fixed to \ensuremath{(0,0)}). \ensuremath{(C_i x, C_i y)} are the current x-/y-coordinates of node \textit{i}. Note that unlike in the literature \cite{43,44},
we do not require the number of nodes to be constant over time.
\subsubsection{Modelling the cumulative distance traveled (\texttt{CDT})}\label{sec:evaluation:DistanceTraveled:modelling:CDT}
The \textit{cumulative distance traveled} (\texttt{CDT}) is measured recursively over the iterations the network has undergone.
At time \ensuremath{t_0} the combined movement of all cells is zero, at any other time \ensuremath{t_{k}} the \texttt{CDT} is the sum of all \texttt{CDT}s in preceding time-steps plus the movement since \ensuremath{t_{k-1}}:

\begin{equation*}
\texttt{CDT}(\ensuremath{t_{0}}) = 0
\end{equation*}
\begin{equation*}
\texttt{CDT} (\ensuremath{t_{k}})=
    \frac{1}{n (t)}
    \sum_{i=1}^{n (t)} |d(pos_{t-1}, pos_{t})| +
\texttt{CDT} (\ensuremath{t_{k-1}})
\end{equation*}
With \ensuremath{|d(pos_{t-1}, pos_{t})|} the absolute Euclidean distance between the previous location of a node and its current one.
\subsubsection{Implementation}\label{sec:evaluation:DistanceTraveled:implementation}
\texttt{CDT} is implemented as a variable that is iteratively updated in Alg. \ref{alg}: the value for the distance moved (\ensuremath{d(pos_{t}, pos_{t-1})}) is calculated as follows:
\begin{equation*}
d(pos_{t}, pos_{t-1}) =
    \sqrt{
    (|C_t x - C_{t-1} x|)^{2} +
    (|C_t y - C_{t-1} y|)^{2}}
\end{equation*}

In Alg. \ref{alg}, the value for \ensuremath{d(pos_{t}, pos_{t-1})}) are then stored in the variable \ensuremath{\textit{shift}_{i}}; the calculation of \texttt{CDT} is a question of aggregating these values (in the variable \ensuremath{\textit{shift}}).

With regard to \texttt{ADT}, since all nodes enter at the same injection point \ensuremath{(0,0)}) this can be easily calculated:
\begin{equation*}
\texttt{ADT} =
    \sqrt{
    \left(\sum_{i=1}^{n} C_i x\right)^{2} +
    \left(\sum_{i=1}^{n} C_i y\right)^{2}}
\end{equation*}
\subsection{Deployment behaviour over time (velocity)}\label{sec:evaluation:NodeVelocity}
%
\subsubsection{Motivation}
When evaluating the performance of BISON we are not just interested in the distances traveled by nodes but also in their collective behaviour. In line with \cite{60-n} we will use the \textit{average} node-velocity to characterize \textit{node behaviour} during the deployment of the WSN.

In our simulation, individual node velocity is determined by the calculation of the Voronoi centroid combined with the node's inference of its location (i.e., the \textit{to} and \textit{from} of the node's movement). The way we determine movement defines the overall collective deployment behaviour but considering the distance (as we do with \texttt{ADT} and \texttt{CDT}) only gives us insight into the performance of the entire system and does not tell us anything about the performance of the nodes in the context of the behaviour of the other nodes.
\subsubsection{Modelling node--diffusion and --drift}\label{sec:evaluation:NodeVelocity:DiffusionDrift}
In line with \cite{60-n} we avoid using a theoretical complex model / equation to determine node-velocity; instead we use empirical data (generated by our simulations) to estimate the diffusion (\ensuremath{D}) and the drift (\ensuremath{F}) coefficients of this equation. The former will capture the mean rate of change of average node velocity, while the latter quantifies the evolution thereof.

The coefficients for diffusion (\ensuremath{D}) and drift (\ensuremath{F}) are defined as follows \cite{60-n}:

\begin{equation}\label{eq:diffusion}
D\big(v(t)\big) = \frac{1}{2} \Bigg \langle\frac{\big( v(t + \delta t) -v(t)
\big)^2}{\delta t}\Bigg \rangle
\end{equation}

\begin{equation}\label{eq:drift}
F\big(v(t)\big) = \Bigg \langle\frac{v(t + \delta t) -v(t)}{\delta t}\Bigg \rangle
\end{equation}

with
\begin{equation*}
v(t) = \frac{1}{n} \sum_{i=1}^{n}{v_i(t)}
\end{equation*}
where $v(t)$ is the average velocity of all the available nodes $n$ at time $t$,
$D\big(v(t)\big)$ is the diffusion coefficient,
$F\big(v(t)\big)$ the drift coefficient and $\delta t$ the chosen time step value.
\subsection{Energy cost incurred for deployment}\label{sec:evaluation:EnergyConsumed}
\subsubsection{Motivation}
Nodes in wireless sensor networks almost always operate with a limited energy budget. Since the energy consumption of individual nodes as well as that of the entire WSN are of paramount interest to practitioners in the field, this has to be considered in our evaluation.

Generally speaking, energy efficiency is commonly measured over the lifetime of a network \cite{62,72}:
\begin{equation}
\textit{WSN efficiceny} =\frac{E_{total}}{E_M + E_S + E_{other}}
\end{equation}
where $E_{total}$ is the total energy that sensor nodes are provided with, $E_M$ is the cost incurred by moving the nodes and $E_S$ is the cost incurred by operating the sensing equipment. To account for more complex implementations we can add $E_{other}$, sub-summing any additional power requirements.

\subsubsection{Modelling}\label{sec:evaluation:EnergyConsumed:modelling}
In the context of this article, we are only concerned with the energy cost incurred by the movement of a node \ensuremath{i}. In the literature \cite{56,55},  \ensuremath{E_{M,i}} was modeled (based on kinetic energy) as follows:
\begin{equation}\label{eq:energy}
E_{M_i}=\frac{1}{2}m_{i}v_{i}^{2}=\frac{1}{2}m_{i}{\left(\frac{\Delta d_{i}}{\Delta t_{i}}\right)}^2
\end{equation}
where \textit{m} is the mass of the sensor node, $t_i$ is the time step of node \textit{i}, $d_i$,  is the distance moved during time step $t_i$, and $v_i$ is the corresponding velocity of the node at $t_i$. Under the reasonable simplifying assumption that the work-energy theorem applies, this expression could be reworked into:
\begin{equation*}
E_{M_{i}}=m{\frac{\Delta^{2} d_{i}}{\Delta t_{i}^{2}}}d_{i}
\end{equation*}
We will use mass-normalized energy, \ensuremath{{E_{M, i}}/{m}}, to further focus on kinetic aspect of the problem.
\subsection{Uniformity of the  coverage}\label{sec:evaluation:NodeDistribution}
\subsubsection{Motivation}
As the area in a circle (sphere) increases non-linearly with the radius, so does the signal strength of a broadcast diminish with distance to the emitting device. Providing a certain signal over a distance does therefore incur an energy cost that is also non-linear in the distance to the source. Therefore, and in all generality, if complete coverage can be achieved by all nodes in a network operating at the same coverage radius then this is guaranteed to be (as far as signal cost is concerned) the best possible set-up.

For us this translates to the goal to reduce coverage overlap as much as possible. Since we are operating in an obstacle-rich environment we do not compare the coverage radii but instead the areas inside the Voronoi regions.
\subsubsection{Modelling}\label{sec:evaluation:NodeDistribution:modelling}
Following \cite{45}, we capture the extent to which Voronoi regions covering an area \ensuremath{A} are uniformly distributed over the nodes in a network by measuring the similarities between the generated polygon regions \ensuremath{U_A}:
\begin{equation} \label{eq:energy.coverage}
U_A =\frac{1}{\textit{mean}({A_V})}
\sqrt{\frac{1}{|\textit{N}|} \sum_{n \in N}\big(  A_{V_i}-\textit{mean}({A_{V}})\big)^2}
\end{equation}
where $A_{V_i}$ is the area of Voronoi cell \textit{i} and $N$ is the total number of nodes in the network. The smaller the value of $U_A$  the better the system is towards uniform distribution.
\section{Results and Discussion}\label{sec:results}
We evaluate the performance of BISON over time and under increasing noise-levels. We focus on two criteria: (1) the movement required to continuously grow the network until the termination criteria halt the algorithm and (2) the growth of coverage during deployment. We furthermore investigate the resilience of the network against node loss and provide a comparison to another approach for some insight into the convergence properties of the approach.
\subsection*{Overview}
In \S \ref{subsec:analysis.and.results:MaterialsMethods} we first provide an overview over the implementation choices taken and sum up the various parameter settings discussed earlier in this article. We then report on three investigations into the performance of BISON:
(1) in \S \ref{subsec:results.movement} we evaluate the algorithm with regard to the movement of the nodes in the different environment types we considered. There, we also present the results of comparing BISON to the NSVA algorithm with regard to performance and convergence.
(2) We then (\S \ref{subsec:results.coverage}) discuss the coverage achieved by a WSN deployed through BISON.
(3) Finally, in \S \ref{subsec:results.Swarm.like.noise} we present our findings and considerations with regard to real-world aspects of swarming and node failure.
\subsection{Materials and methods}\label{subsec:analysis.and.results:MaterialsMethods}
The BISON algorithm (provided as Alg. \ref{alg}) was implemented using MATLAB version R2017b. As mentioned, there are a number of control parameters and settings.
Unless otherwise stated the following values were used in simulations for data collection and performance evaluation:

\begin{itemize}
    \item sensing range: \ensuremath{R_S = 1m} (intended interpretation: \ensuremath{m =} one \textit{meter}); communication range: \ensuremath{R_C = \sqrt{3} R_S}.
    \item simulated environment: \ensuremath{10 \times 10 m^2}.
    \item inject node angle \ensuremath{\theta}:  random between \ensuremath{0^\circ} and \ensuremath{90^\circ}.
    \item injection point (aka \textit{origin}): \texttt{(0,0)}.
    \item termination criteria:
        shifting threshold \ensuremath{\tau = \frac{R_S}{100}} (\S \ref{subsec:BISON.node.movement});
        rounds without improvement \texttt{c\ensuremath{_{max}}}\ensuremath{ = 15} (\S \ref{subsec:Model.noise.2nd.stop}).
\end{itemize}

The following values were recorded during simulations:
\begin{itemize}
    \item \underline{movement} (Euc. dist.) of any node in the WSN \ensuremath{i}
        (\ensuremath{shift_i}, \S \ref{sec:evaluation:DistanceTraveled:implementation}) to assess
        average 
    (\texttt{ADT}, \S \ref{sec:evaluation:DistanceTraveled:modelling:ADT}) and
    cumulative (\texttt{CDT}, \S \ref{sec:evaluation:DistanceTraveled:modelling:CDT}) distance traveled
    and to calculate the movement energy cost (\ensuremath{E_{M}}, \S \ref{sec:evaluation:EnergyConsumed:modelling}).
    \item \underline{area covered}: absolute
        (\S \ref{subsec:BISON.node.movement})
        and in \%: \texttt{PAC}
        (\S \ref{sec:evaluation:PAC:modelling}) as well as the uniformity of the coverage  (\ensuremath{U_A}, \S \ref{sec:evaluation:NodeDistribution:modelling}).
\end{itemize}

\subsection{Performance evaluation: movement}\label{subsec:results.movement}
\subsubsection{Movement in obstacle-free environments}\label{subsec:results.movement:obstacle.free}
\begin{figure}[h!]
    \centering
    \includegraphics[width=\hsize]{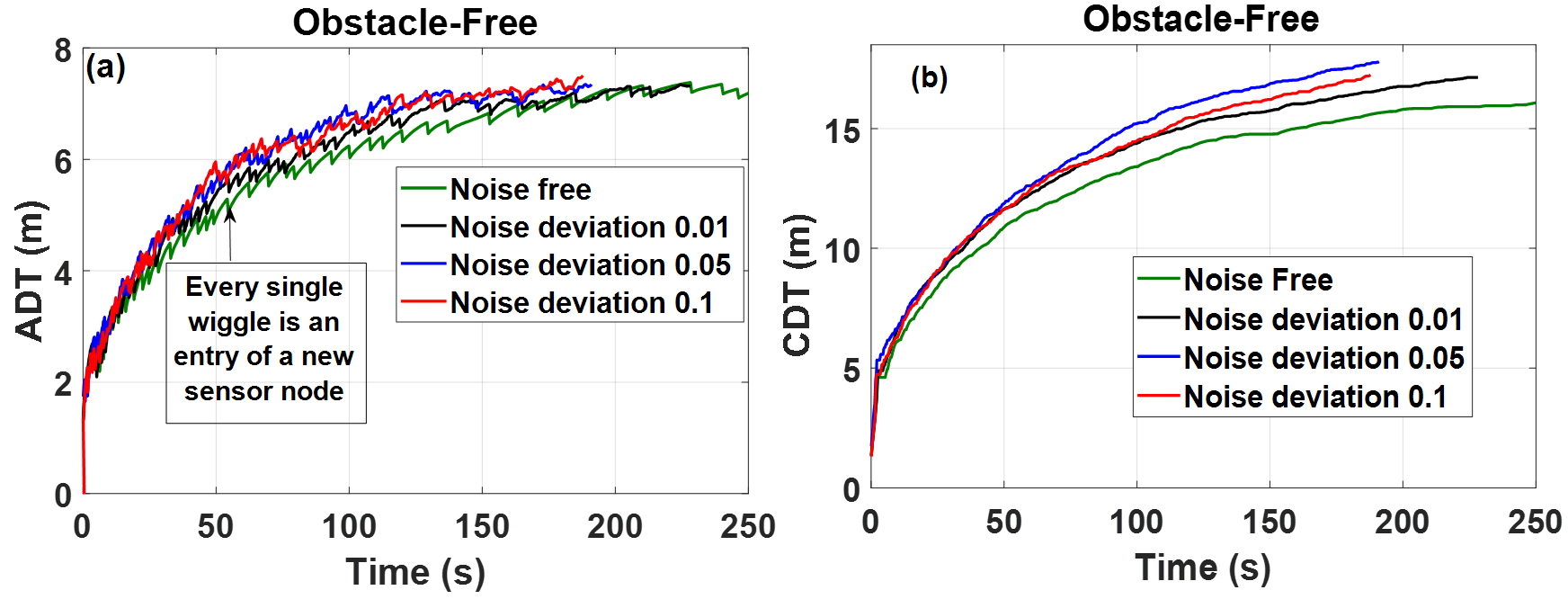}
    \caption{Initial investigation into the movement to deploy a WSN into an obstacle-free environment. Notably, the highest noise level (\ensuremath{\sigma = 0.1}) outperforms the second highest (\ensuremath{\sigma = 0.05}) in distance traveled. Furthermore, increasing noise consistently improves deployment time! This, initially counter-intuitive, result is discussed separately in \S \ref{subsec:results.swarming:impact.of.noise}.}
    \label{adt-obf}
    \end{figure}

To establish a baseline, BISON was used to deploy a WSN into an obstacle-free environment while the nodes were subjected to various levels of noise. Both, the average- (\texttt{ADT}) as well as the cumulative- distance traveled (\texttt{CDT}) by the nodes in the WSN were recorded and are plotted in Fig. \ref{adt-obf}.

We would expect that the performance suffers as noise-levels increase. Indeed, at least with regard to the distances traveled this is initially the case for both \texttt{ADT} and \texttt{CDT}. However, upon closer inspection we see that for both measures the highest noise-level actually outperforms the second most severe noise-level: for \ensuremath{\sigma = 0.1} the network requires less movement or time than it does for \ensuremath{\sigma = 0.05}.

    \begin{figure*}[htbp]
    \centerline{\includegraphics[width=0.52\hsize]{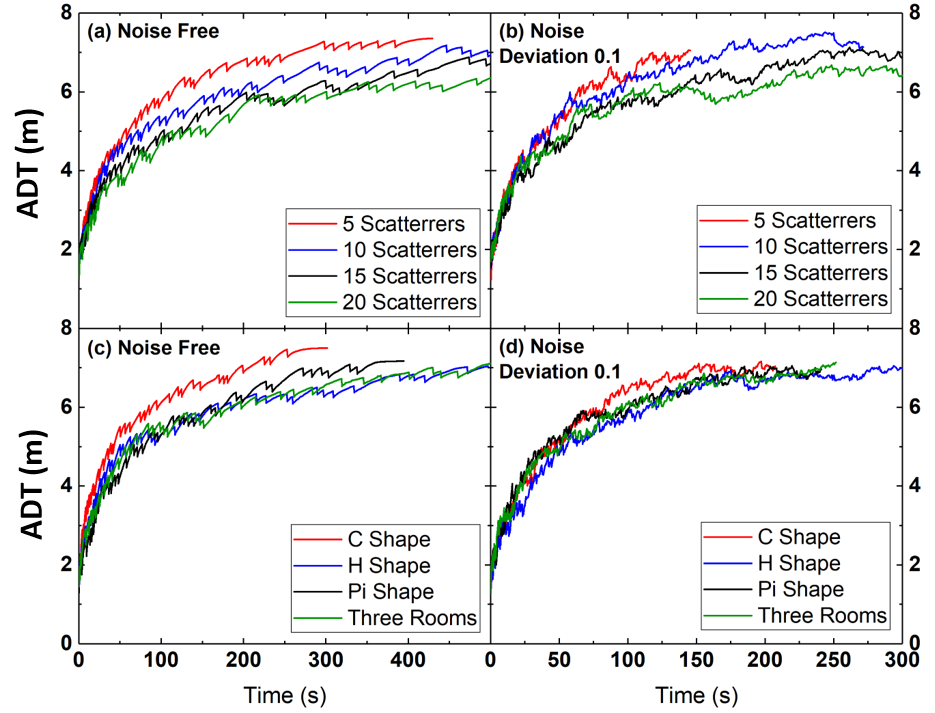}
                \includegraphics[width=0.55\hsize]{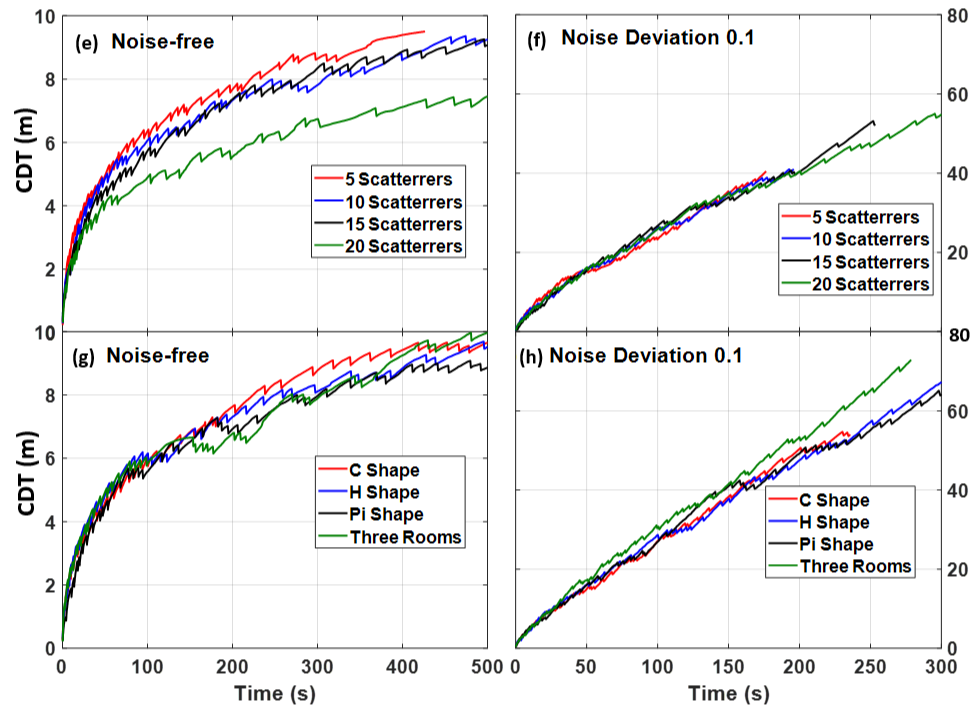}}
    \caption{The average- (\textnormal{\texttt{ACT}}, left 4 plots) and cumulative- (\textnormal{\texttt{CDT}}, right 4 plots) distance travelled by nodes in obstacle-rich environments (top row: pillars scattering the signal; bottom row: different wall configurations blocking the signal) and the impact of noise (respective right columns). Consistent with the results for obstacle-free environments, shown in Fig. \ref{adt-obf}, noise consistently improves deployment speed (measured in time, x-axis) but does so at significant cost in cumulative distance traveled to reach the final deployment position (cf. rightmost two plots, see y-axis).}
    \label{adt-5.cdt-5.combined}
    \end{figure*}

Furthermore, for both measures, the time required to reach the stop condition
consistently \textit{decreases} as the noise-level increases. This implies that noise can have a beneficial impact on collective performance.
These results are consistent with the literature (cf. \cite{45} on the NSVA algorithm in noisy environments); \S \ref{subsec:results.Swarm.like.noise} is dedicated to exploring this, counter-intuitive result in more detail.

The \texttt{CDT} turns out to be roughly twice the \texttt{ADT}. As we will continue to compare these two measures it seems important to point out what exactly they represent again: both report an average value, but \texttt{ADT} uses the \textit{as-the-crow-flies} distance while \texttt{CDT} uses the actual distance traveled. Thus, the factor by which \texttt{CDT} exceeds \texttt{ADT} is an indicator for the \textit{detours} taken by the nodes to reach a position.
\subsubsection{Movement in obstacle-rich environments}\label{subsec:results.movement:obstacle.rich}
The positive effect of high (\ensuremath{\sigma = 0.1}) noise-levels - noticed in the previous section - is also consistently found in the results presented in this section. To enhance readability of  Fig. \ref{adt-5.cdt-5.combined}, we only report on the outcomes for \ensuremath{\sigma = 0.1} and \textit{``no noise at all''};
the interested reader is referred to Figs. S4 and S5 in SOM for the other results, omitted here.

For obstacle-rich environments we report \texttt{ADT} and \texttt{CDT} separately again (cf. Fig. \ref{adt-5.cdt-5.combined}). We compare performance under lowest and highest noise level and for environments with signal scattereres (i.e., pillars) or walls.
\subsubsection*{The impact of noise}
The first thing that we notice when comparing the results plotted in Fig. \ref{adt-5.cdt-5.combined} is that for \texttt{ADT} and \texttt{CDT} as well as for both types of environments, the algorithms achieves coverage \textit{faster} in the presence of noise.
\subsubsection*{Walled / scattered environments}
Increasing the number of signal scattering objects in the environment reduces the distance traveled by the nodes, especially for \texttt{ADT}.

While increasing the noise-level actually improve collective performance (cf. \S \ref{subsec:results.Swarm.like.noise}) the reduced distance here can be - at least partly - explained by the fact that there is effectively less of the environment to cover as an increasing percentage thereof is filled by the pillars (cf. Fig. \ref{obstacles}). However, for the 4 walled experiments this explanation does not hold: the scenarios \textit{C-shaped} and \textit{3 rooms} have significantly less space taken up by their respective walls than the other two environments (cf. Fig. \ref{shapes}), yet their \texttt{ADT} differs greatly. This is not unexpected because, especially in case of \textit{3 rooms}, there is a bottleneck for the nodes to pass through. This is supported by the corresponding \texttt{CDT} where this scenario requires the largest cumulative distance traveled.

\subsubsection*{Different variations on environment types}
The performance difference for different cases of an environment type was consistent across experiments, indicating that some environments are inherently more difficult to solve than others. An theoretical study of which scenario aspects this depends on is outside the scope of this paper.
    \begin{figure}[h!]
    \centerline{\includegraphics[width=\hsize]{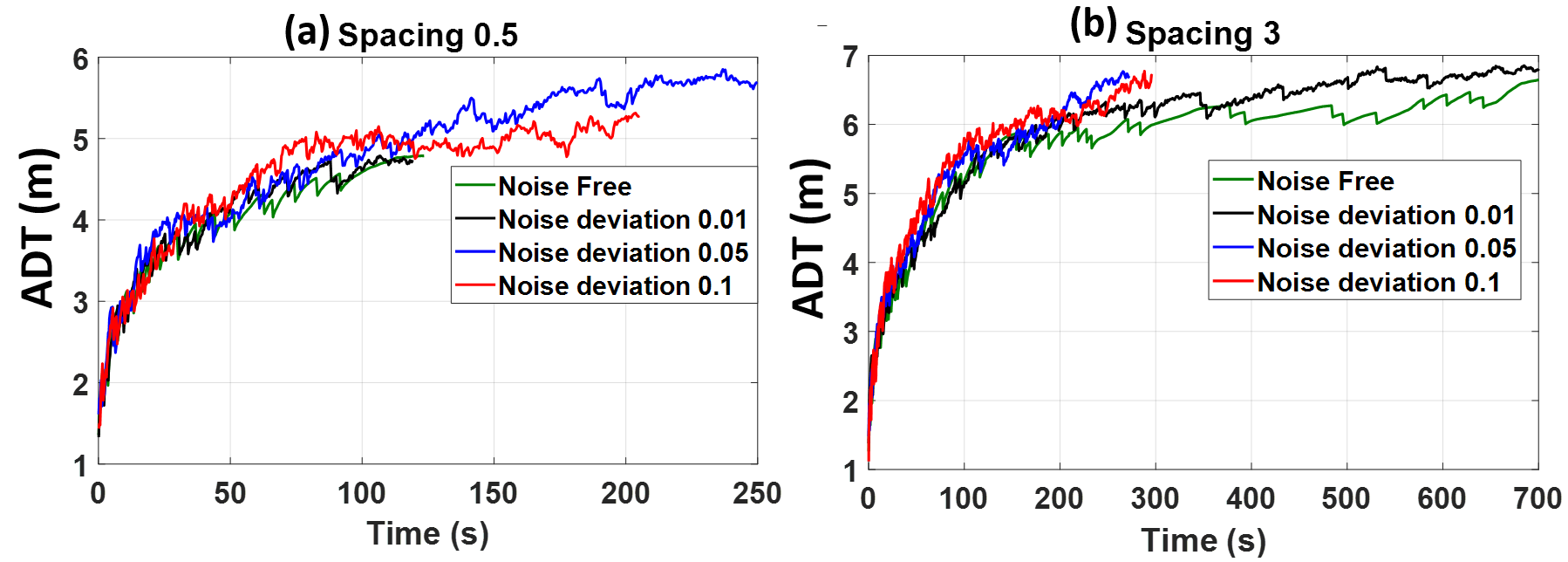}}
    \caption{\textnormal{\texttt{ADT}} performance for environments with small spaces of different width (see Fig. \ref{spacing}). Plotted are the smallest (a) and widest (b) of these scenarios. Note:  graphs shown in (a) are from a deployment where the network could not actually cover the cavities. As before (cf. Figs. \ref{adt-obf}, \ref{adt-5.cdt-5.combined}), adding noise significantly improves performance.}
    \label{adt-spacing}
    \end{figure}

    \begin{figure*}[h]
    \centerline{\includegraphics[width=0.2\hsize]{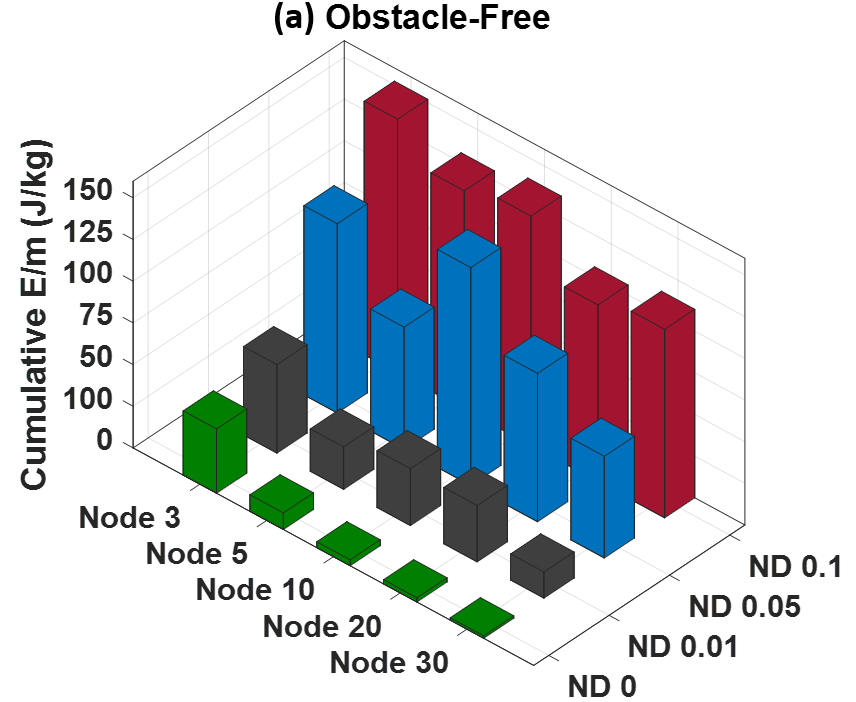}
                \includegraphics[width=0.4\hsize]{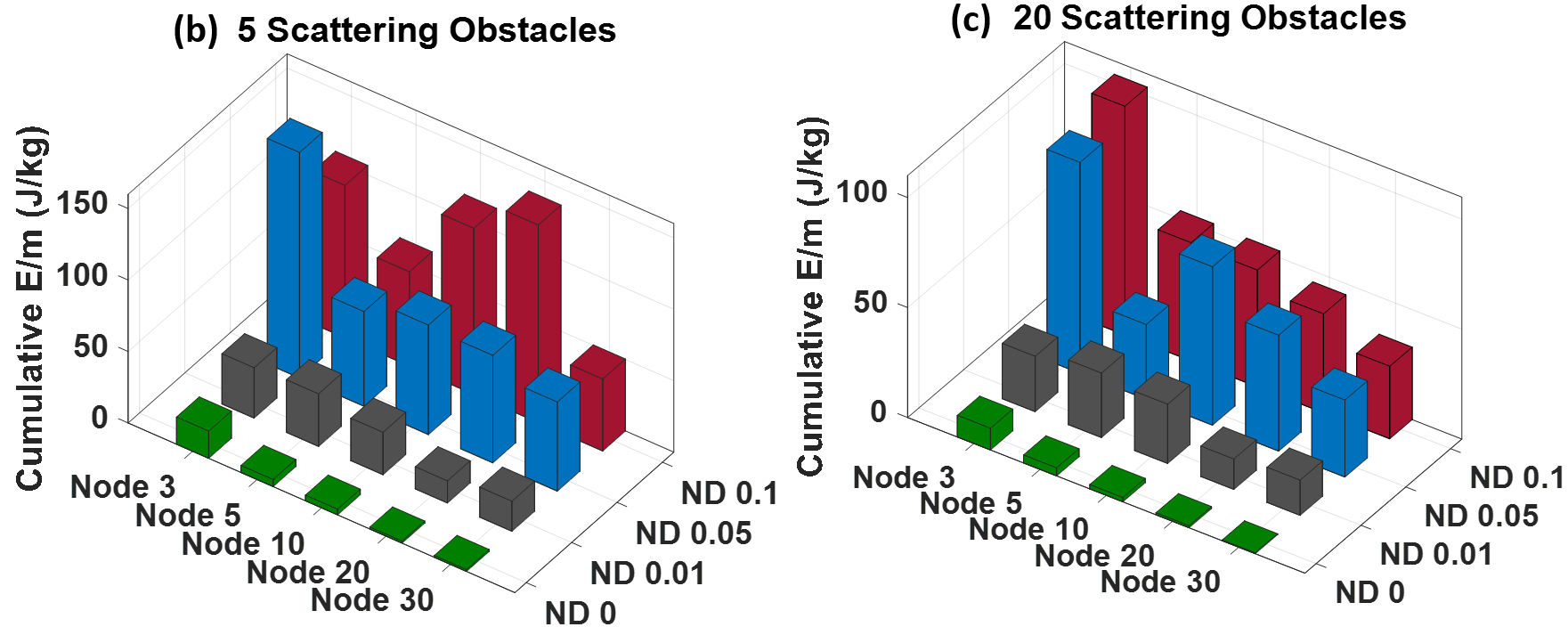}
                \includegraphics[width=0.4\hsize]{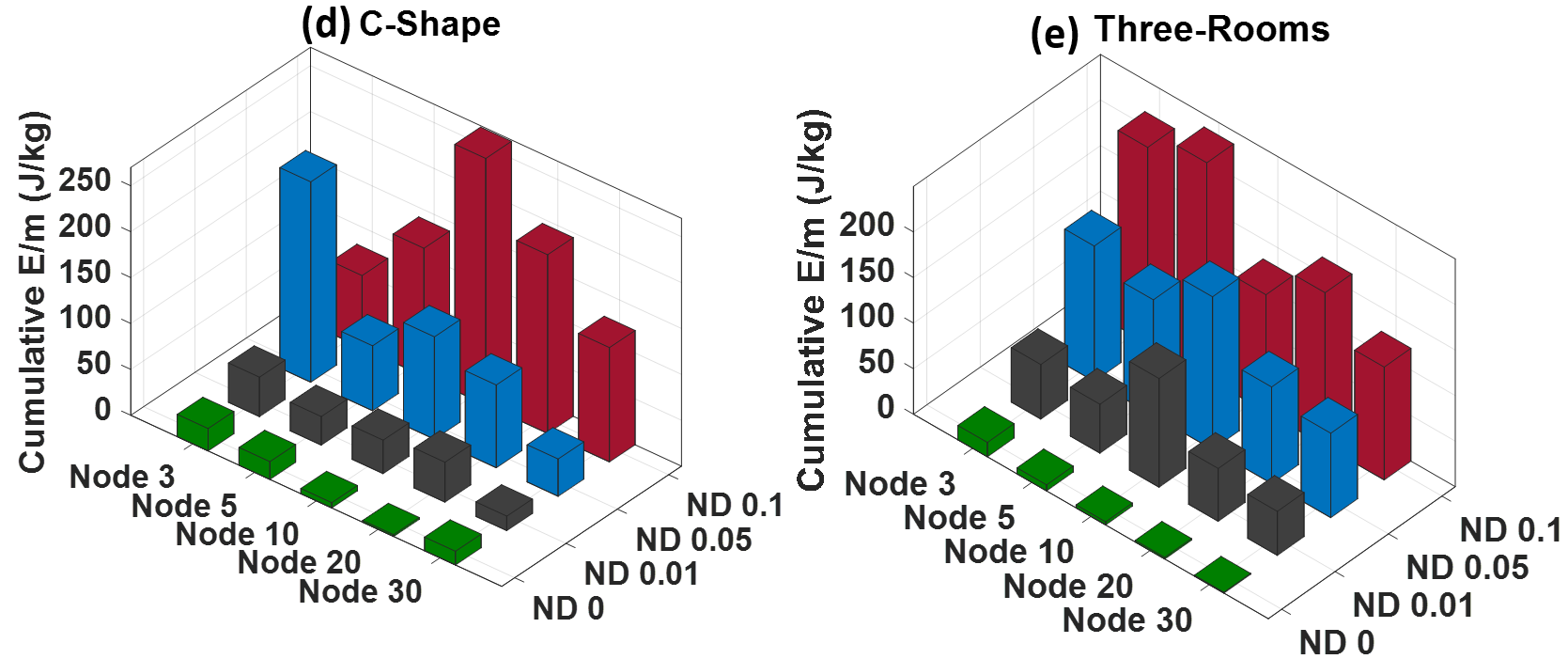}}
    \caption{Representation of individual cost for arbitrarily selected nodes (the number indicates the order of injection with lower nodes having spent more time in the environment) in different environments. The improvement in deployment speed (caused by increased noise-levels) is thus achieved at the cost of significantly increased energy consumption (cf. Equation \ref{eq:energy}). Environments differ only in the type and location of obstacles. The impact thereof is clearly visible when comparing the results, though the relation between placement and energy cost is not immediately obvious.}
    \label{energy.combined}
    \end{figure*}
\subsubsection{Movement in environments with crevices}\label{subsec:results.movement:obstacle.crevices}
First of all we refer back to Fig. \ref{spacing}, plot (a), to remind the reader that in the most restrictive scenarios (those where the spaces are merely a fourth of the nodes' sensing range) the network simply fails to extend into these crevices. This obviously results in sub-optimal coverage but also means that less nodes are needed.  Note that for all the tested scenarios (including the empty one), this was the only one where no noise, or low noise-levels outperformed higher noise-levels (cf. the plots in Figs. \ref{adt-obf}, \ref{adt-5.cdt-5.combined} and \ref{adt-spacing}).
Avoiding the openings altogether does, however, impact the performance because the nodes facing these openings should be subjected to small incentives to move towards the openings (or into them) as their Voronoi regions are not restricted in that direction.

We report only on two of the 4 scenarios here; additional results, including results for all simulated scenarios and noise-levels are provided in S6 in SOM.

\subsubsection{Cost incurred through node movement}\label{subsec:results.movement:energy.cost}
Looking back at the plotted \texttt{CDT} values in Fig. \ref{adt-5.cdt-5.combined}, we clearly see that while the addition of noise consistently resulted in improved performance \textit{time} it did so at a significant cost through increase in movement, as evident from Fig. \ref{energy.combined}.
\pagebreak

\begin{figure}[htbp]
    \centerline{\includegraphics[width=\hsize]{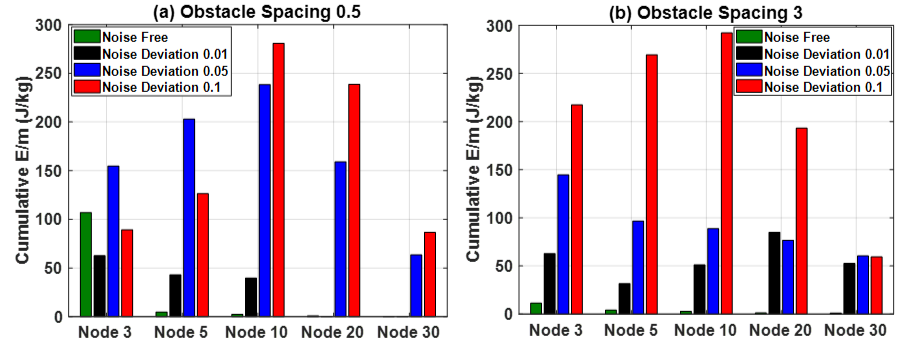}}
    \caption{The nodes specific energy cost for deployment into large environments with small cavities, visualized in Fig. \ref{adt-spacing}.}
    \label{work-spacing}
    \vskip-1ex
    \end{figure}

When comparing the individual node movement with the resulting energy cost (cf. Figs. \ref{adt-obf} and \ref{energy.combined} for obstacle free-- and Figs. \ref{adt-5.cdt-5.combined} and \ref{energy.combined} for obstacle-rich -- environments; cf. Figs. \ref{adt-spacing} and \ref{work-spacing} for environments with small spaces) we consistently see the trade-off.
While the plots provided (reporting on the energy expenditures of \textit{some} of the nodes) suffice to show that it is the environment and not the order of injection that determines the final energy cost,
the full set of results / additional energy representation for obstacle-rich environments and cavities are provided on the SOM.

\subsubsection{Convergence properties of BISON}\label{subsec:results.movement:convergence}
\begin{figure}[h!]
    \centerline{\includegraphics[width=\hsize]{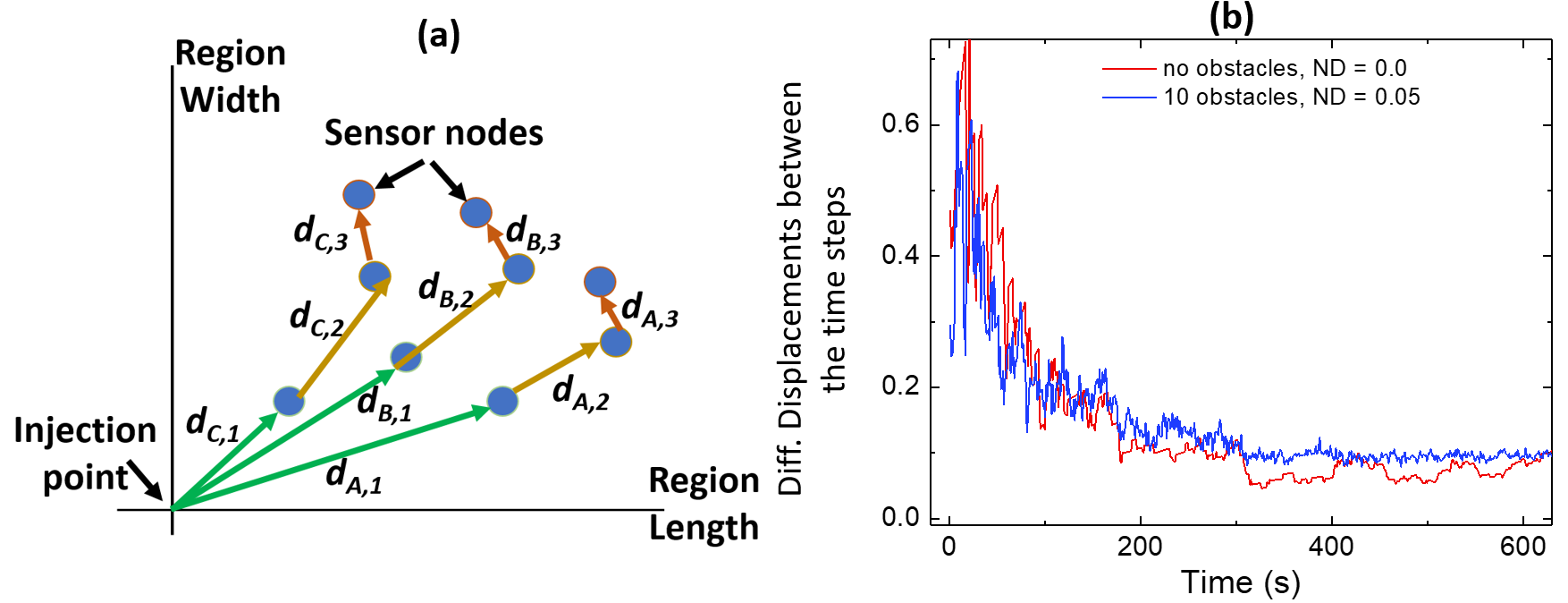}}
    \caption{(a) shows the distance travelled for nodes over time, with \ensuremath{d_{A,1}} the distance moved by node \ensuremath{A} in time step \ensuremath{1}. In (b) we plot the Differential Displacements (step-to-step displacements) as a function of time. After around \ensuremath{t=200} the observed movement is entirely caused nodes reacting to other node's movement as coverage is achieved.}\label{convergence}
    \end{figure}

We find no evidence that the modifications we introduced 
fundamentally alter the arguments (presented elsewhere in the literature \cite{vbook} in favor of convergence of Voronoi algorithms. Nevertheless, we conducted a numerical experiment to compare performances of BISON  
(cf. Fig. \ref{convergence}).
\subsubsection*{Benchmarking of results}
We compared node movement before and after coverage was achieved. Perhaps unsurprisingly, we see a rapid rise of the differential displacement as a function of time. After a maximum is reached, the motion of sensor nodes in the system rapidly decreases (cf.  Fig. \ref{convergence}(b)). The maximum occurs close to the instant in time where a derivative of the \texttt{PAC} and \texttt{ADT} curves occur.

We furthermore compared BISON to closest related algorithm in the literature: the NSVA algorithm \cite{43,44}. The choice for NSVA was made due to its similarity in terms of the assumptions made, parameters used, as well as simulation time steps. In addition, it was important to use to compare to an approach that generates the Voronoi regions based on local information about the environment from the node’s sensing range, where other related algorithms (e.g., \cite{40,41,38,39,42a,46a,46b}) use global information or apply to heterogeneous nodes, which require different forms of Voronoi regions altogether.
Finally, since the metrics used are, in fact, proposed in \cite{43,44,45}, the NSVA algorithm is the most obvious candidate for a comparative evaluation.

\begin{table}[h!]
\centering
 \caption{Representation of the parameters used to compare BISON with NSVA algorithm and the resulted PAC and ADT}
 \label{table1}
\renewcommand{\arraystretch}{1.5}
\begin{tabular}{ | c | c | c |  }
\hline
Quantity & NSVA & BISON\\
\hline\hline
Area (\ensuremath{m^2}) & $100 \times 100$ & $100 \times 100$\\
\hline
\ensuremath{R_S} (\ensuremath{m}) & 16 & 16\\
\hline
\ensuremath{R_C} (\ensuremath{m}) & 16 & $\sqrt{3}\times 16$\\
\hline
Number of nodes (\ensuremath{n}) & 40 & 16\\
\hline
Time (\ensuremath{s}) & 150s & 20s\\
\hline
\texttt{PAC} (\%) & 93.4\% & 99.3\%\\
\hline
\texttt{ADT} (\ensuremath{m}) & 74 & 85\\
\hline
\end{tabular}
\end{table}

\begin{figure}[h!]
    \includegraphics[width=\hsize]{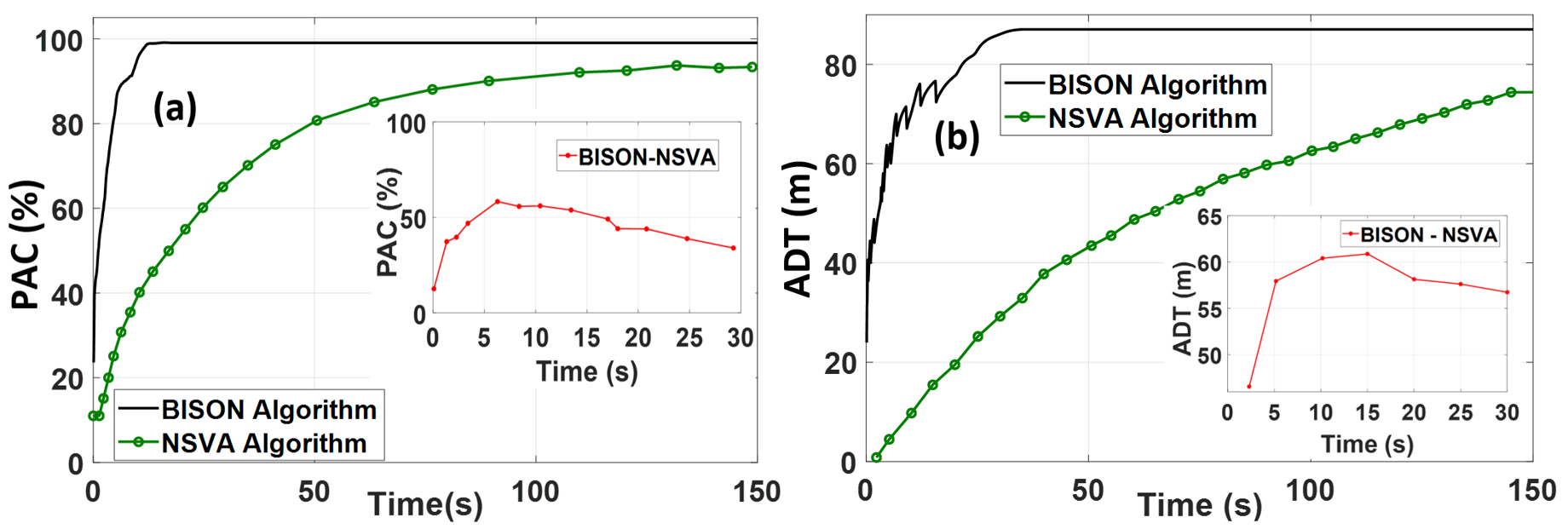}
    \caption{Comparing the performance of BISON and NSVA over time: for both, \textnormal{\texttt{PAC}} (left) and \textnormal{\texttt{ADT}} (right), BISON outperforms NSVA significantly. Since BISON plateaus after around 30\ensuremath{s}, the inserted plots visualize the difference in respective performance for that time.}
    \label{nb.combined}
    \end{figure}

Table \ref{table1} above provides a quick overview over the performance of both algorithms in an obstacle-free environment with no noise effect. The results of our comparative evaluation over time are plotted in Fig. \ref{nb.combined}: the left panel compares the algorithms on the basis of \texttt{PAC} while the right shows the respective \texttt{ADT}. Both inserts plot the difference between BISON and NSVA for the first 30 seconds of the simulation. BISON clearly outperforms NSVA.
\pagebreak

The performance is compared with BISON algorithm under the same parameters specified in Table \ref{table1}. It can be noticed that NSVA algorithm required 40 nodes and 150\ensuremath{s} to achieve 93.4\% of area coverage, while BISON algorithm required one fifth of that time (terminating after \ensuremath{\approx} 30\ensuremath{s}) with 16 nodes to achieve 99\% of area coverage. The improved speed can partly be attributed to the wider communication range \ensuremath{R_C}, but this does not explain why BISON can operate with a significant smaller  number of nodes.

To be fair, the final coverage is far more indicative of the superiority of BISON than the progress over time. The reason for this is that NSVA injects a fixed number of nodes inside the environment \cite{43,44} (here: 40) at start time while BISON injects its node as needed over time. Due to this, the plotted \texttt{ADT} in Fig. \ref{nb.combined}, right, has to be much better for BISON as most nodes in NSVA cannot move initially. However, BISON achieves a better coverage faster (cf. Fig. \ref{nb.combined}, left), and does so using significantly fewer nodes. In addition, lower energy consumption suggests a longer lifetime of the BISON network.

The observed difference between BISON and NSVA, shown in Fig. \ref{nb.combined}, inserts (a), (b), is only quantitatively different for the various environments and noise levels. 
\subsection{Performance evaluation: coverage}\label{subsec:results.coverage}
We evaluate performance using \textit{percentage area coverage} (\texttt{PAC}), introduced in \S \ref{sec:evaluation:PAC:modelling}.
Analogous to the results for movement (and thus energy cost) we find that increasing noise-levels have a significant effect on the performance: as before, the time required to achieve coverage decreases as noise-levels increase (cf. insert in Fig. \ref{pac-obf}). This comes at the cost of (significantly) increased energy consumption.
\subsubsection{Coverage in obstacle-free environments}\label{subsec:results.coverage:obstacle.free}
\begin{figure}[h!]
    \centerline{\includegraphics[width=0.7\hsize]{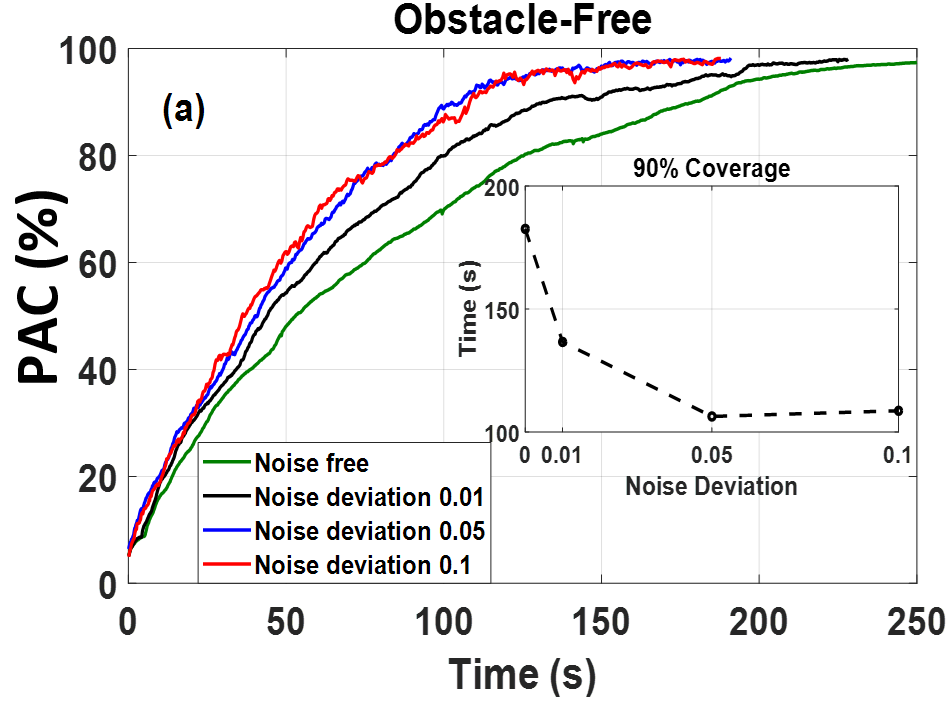}}
    \caption{Performance measured in \textnormal{\texttt{PAC}} in an obstacle-free environment. 
    As before we notice a significant improvement in time with increasing noise, shown also in the inserted graph plotting the time required to achieve 90\% coverage for the different noise-level}
    \label{pac-obf}
    \end{figure}

In obstacle free environments nodes can freely diffuse into the environment, obstructed only by the walls surrounding the scenario. With increasing noise-levels, however, the node's Voronoi regions incur increasing error which directly translates to spotty coverage, affecting the \texttt{PAC}. As shown in Fig. \ref{pac-obf}, the ultimate outcome is nearly the same (differing by less than 1\% in the achieved final coverage, which was 99\% for noise-free environments and 98\% for environments with high noise). However, when looking at the graphs plotted we see a noticeable difference in smoothness, representing the aforementioned spotty coverage for high noise-levels.

\subsubsection{Coverage in obstacle-rich environments}\label{subsec:results.coverage:obstacle.rich}
In the interest of readability, we again restrict ourselves to reporting the results obtained from running the simulations using both extremes for noise-levels, with the remaining results provided in Figs. S10 and S11 in SOM for brevity.

    \begin{figure}[h!]
    \centerline{\includegraphics[width=\hsize]{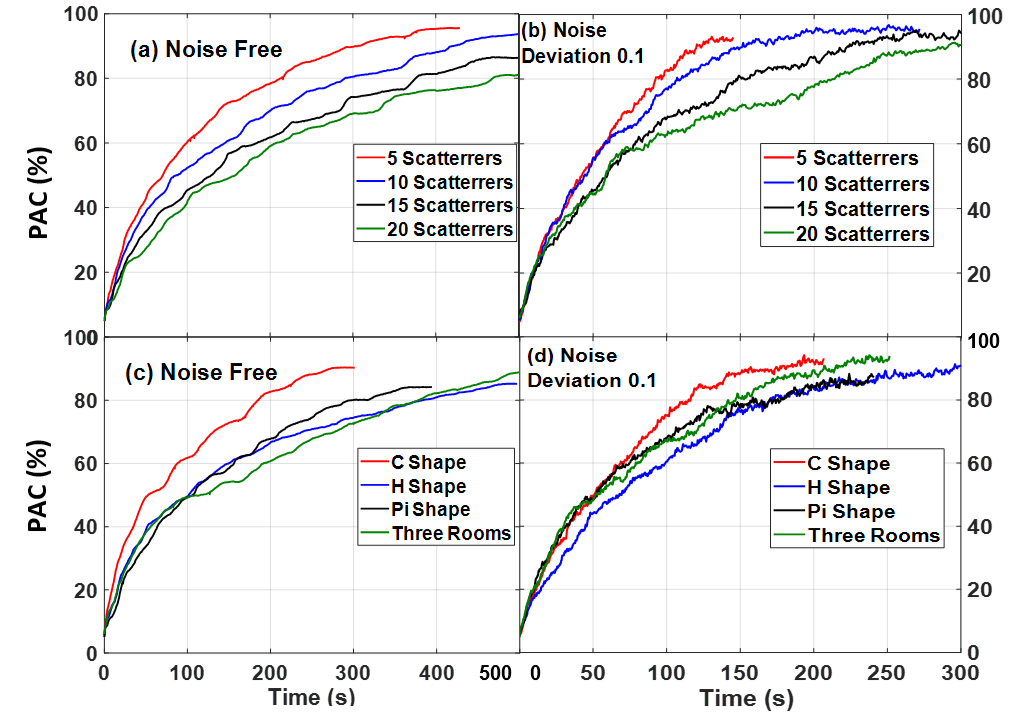}}
    \caption{The \textnormal{\texttt{PAC}} achieved by BISON in obstacle-rich environments with pillars, top (cf. Fig. \ref{obstacles}) or walls, bottom (cf. Fig. \ref{shapes}), plotted over time (x-axis) and \textnormal{\texttt{PAC}} (y-axis). These results correlate to the average- as well as cumulative- distance traveled, plotted in Fig. \ref{adt-5.cdt-5.combined}.}
    \label{pac-5}
    \end{figure}

We considered environments with scatterrers (cf. Fig. \ref{obstacles}),
and walls (cf. Fig. \ref{shapes}), where Fig. \ref{pac-5} plots the \texttt{PAC} under the two most extreme noise-levels. As discussed before, high noise-levels tend improve the speed of the algorithm. The range of \texttt{PAC}  reached by sensor nodes at different noise levels is within $\pm$ 5.0\% (on the absolute coverage scale) between minimum and maximum values reached, demonstrating that BISON can handle noisy environment without dropping the coverage quality, all that while maintaining connected network.
\subsubsection{Coverage in environments with crevices}\label{subsec:results.coverage:obstacle.crevices}
By exploring performance in environments with narrow regions (depicted in Fig. \ref{spacing}) we address the issue of providing network coverage in difficult environments.

\begin{figure}[h!]
    \centerline{\includegraphics[width=\hsize]{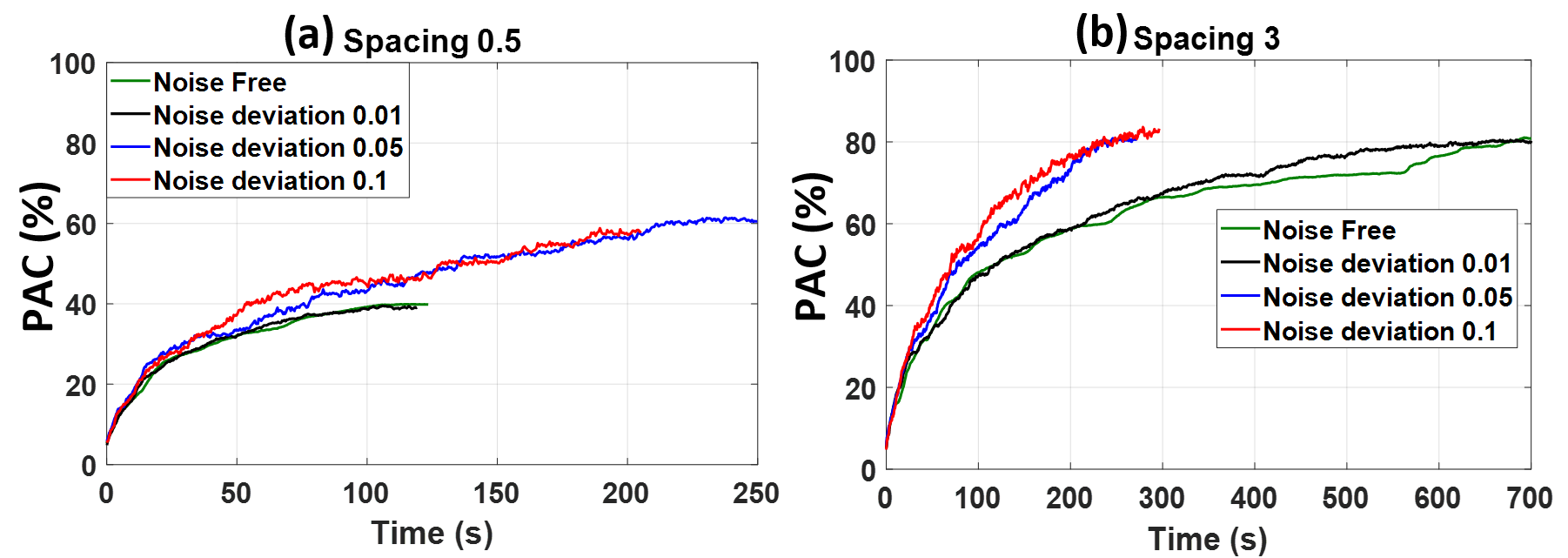}}
    \caption{The impact of noise with regard to \textnormal{\texttt{PAC}} in environments with elongated cavities (see Fig. \ref{spacing}) of width 0.5\ensuremath{m} and 3.0\ensuremath{m}. The 20\% difference in \textnormal{\texttt{PAC}} in scenario (a) is due to the fact that without noise the network never manages to extend into the spacing at all, cf. Fig. \ref{spacing}.}
    \label{pac-spacing}
    \vskip-1ex
    \end{figure}

Fig. \ref{pac-spacing} plots the \texttt{PAC} obtained in these scenarios for spacing sized at 0.5\ensuremath{m} (narrow passage) and  3.0\ensuremath{m} (wide passage), cf. Fig. S12 in SOM for the remaining results.

For narrow spacing (e.g., 0.5\ensuremath{m}), the presence of high noise deviations allows nodes to discover narrow areas due to substantial changes in the Voronoi regions and corresponding centroids generated, thus enabling nodes to cover more regions compared to noise-free or low noise levels.
\subsubsection{Distribution of coverage uniformity}\label{subsec:results.coverage:distribution}
Above we compared performance during deployment using the distance traveled (cf. \S \ref{subsec:results.movement:energy.cost}), which we translated into energy cost using the measure of kinetic energy (cf. Equation \ref{eq:energy}). Analogously, we do the same for the coverage provided, with the energy cost being encoded in the measure of coverage uniformity \ensuremath{U_A} (cf. Equation \ref{eq:energy.coverage}), for which the smaller the value the better the performance.
    \begin{figure}[h!]
    \centerline{\includegraphics[width=\hsize]{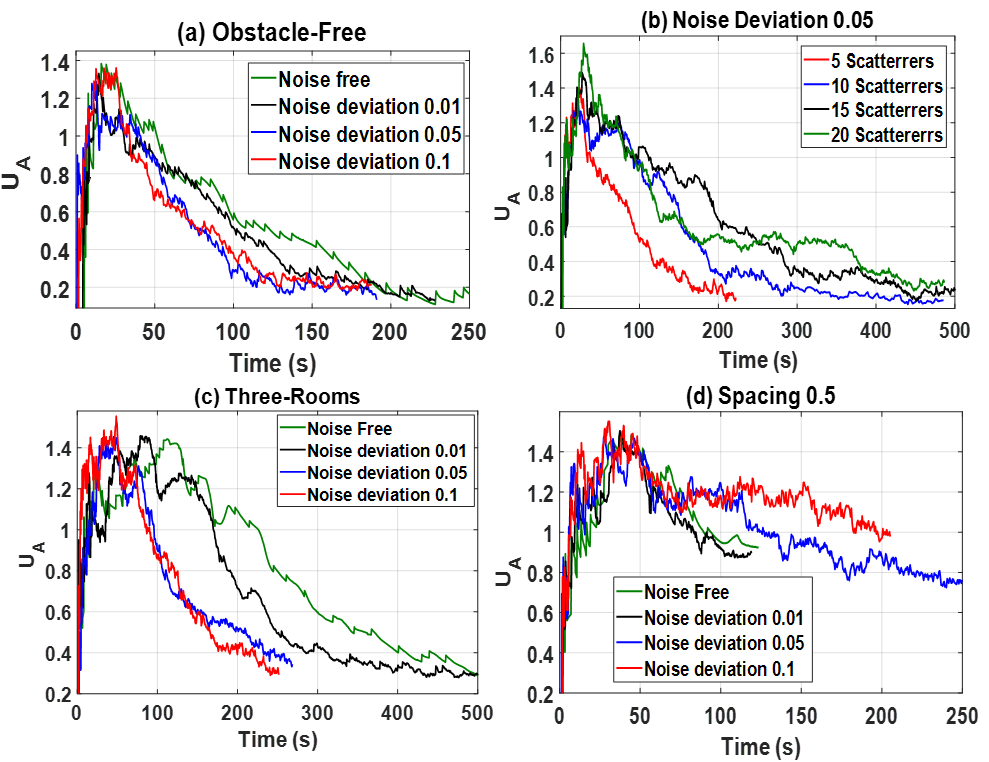}}
    \caption{$U_A$ (cf. \S \ref{sec:evaluation:NodeDistribution:modelling}), lower values represent better performance. (a) and (c) show faster convergence for higher noise-levels, while (d) (small cavities) does not. The plots for various scenarios in (b) confirm that increasing numbers of pillars result in slower convergence of $U_A$.}\label{fig:ua}
    \end{figure}
\subsubsection*{The impact of noise on \ensuremath{U_A}}
With regard to the general impact of noise on the evolution of \ensuremath{U_A}, the results in Fig. \ref{fig:ua} (specifically, plots (a) and (b)) show that increasing noise levels drive faster convergence towards low values for \ensuremath{U_A}. This is explored in more detail in \S \ref{subsec:results.swarming:impact.of.noise}.
\subsubsection*{The impact of obstacles on \ensuremath{U_A}}
Given the expected outcomes for the impact of increasing noise-levels, we investigate the impact of obstacles, Fig. \ref{fig:ua}, (c). Pillars obstruct deployment and impact coverage regions. As expected, increasing the number of obstacles slows down the convergence towards similar coverage. The final results are, however, comparable and differ only marginally. Additional results of $U_A$ for obstacle-rich environments are found in Figs. S13 and S14 in SOM.
\subsubsection*{\ensuremath{U_A} for small elongated spaces}
When deploying nodes into narrow spaces, their Voronoi regions are defined by the walls on either side, causing the final \ensuremath{U_A} to be much higher; this is clearly visible in Fig. \ref{fig:ua}, (d). Furthermore, in the absence of noise cavities are not entered, allowing the WSN to deploy faster into the effectively smaller environment. More illustrative results are found in Fig. S15 in SOM.
\subsection{Swarm-like Noise Coherence in BISON}\label{subsec:results.Swarm.like.noise}
Throughout the paper we have noticed a - somewhat counter-intuitive - impact of noise: increasing noise-levels resulted in better coverage and faster convergence (though at the cost of significantly larger distances traveled).
We argue that this is due to noise-driven coherence in the collective motion of the WSN, which can be seen as a swarm of sensors with Voronoi-distributed domains.

To test this hypothesis, we analyzed the distribution of the diffusion- and drift-coefficients values (cf. \cite{60-n}) for the sensor nodes as a function of the average velocity.
\subsubsection{The impact of noise on performance}\label{subsec:results.swarming:impact.of.noise}
\begin{figure}[h!]
    \centerline{\includegraphics[width=\hsize]{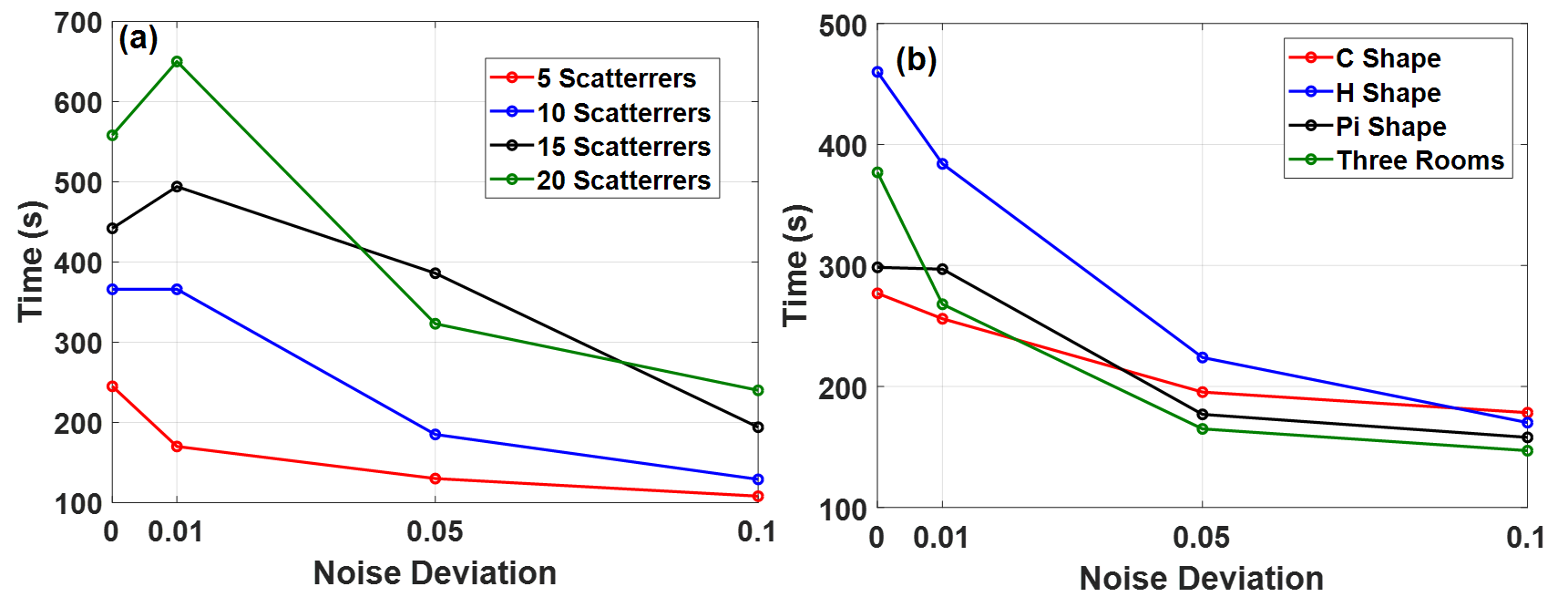}}
    \caption{The impact of increasing noise-levels for environments with pillars (left) and walls (right). Reported values (y-axis) represent achieving 85\% coverage. The impact of noise is undeniable.}
    \label{time-noise}
    \end{figure}

First of all, we confirm the consistent positive impact observed throughout the results: Fig. \ref{time-noise} shows this for our simulated environments with pillars (left) and walls (right), measuring the time to achieve 85\% coverage, over the 4 noise-levels used.

Upon reflection, the explanation for these results is actually straight-forward: increased noise results in increasingly erroneous node localization and, thus, centroid calculation. This in turn leads to increased movement to counter the perceived off-set, which constitutes a faster space exploration.
    \begin{figure*}[h!]
    \centerline{\includegraphics[width=0.5\hsize]{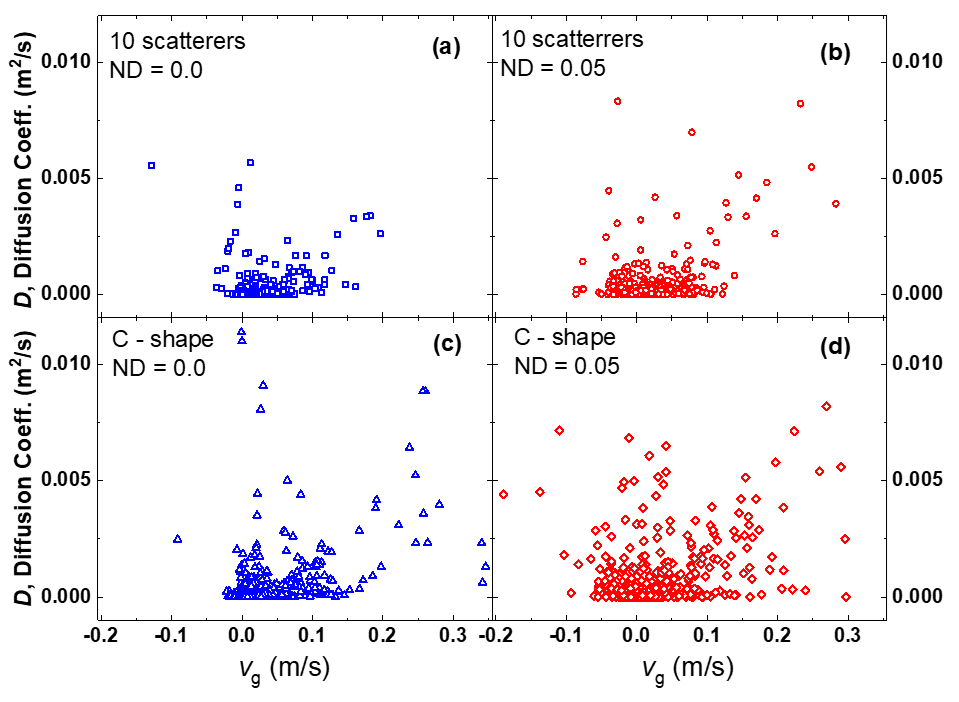}
                \includegraphics[width=0.5\hsize]{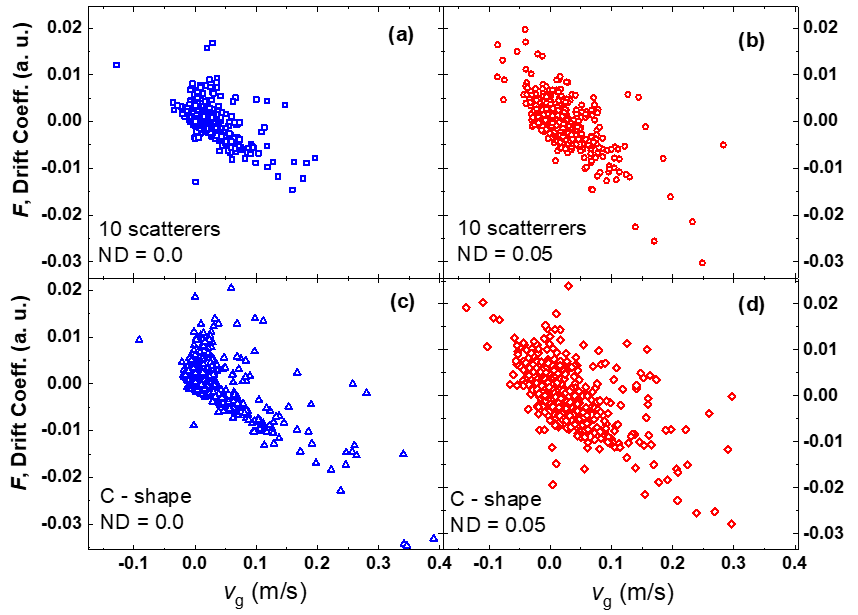}}
    \caption{Investigating collective movement hehaviour on the basis of \textit{diffusion} \ensuremath{D} and \textit{drift} \ensuremath{F} (cf. \S \ref{sec:evaluation:NodeVelocity}). On the x-axis we range over velocity while the y-axes are the values for diffusion or drift, respectively. Since the only variable difference between the respective left (blue) and right (red) panels is the noise-level (\textit{``no noise''} on the left in blue, noise-level \ensuremath{\sigma = 0.05} on the right in red), it is clear that the motion of the sensors is affected by noise.}
    \label{coefficient.combined}
    \end{figure*}
\subsubsection{The impact of noise on collective motion}\label{subsec:results.swarming:colelctive.motion}
Collective motion of objects of both, biological and artificial nature, may correspond to various forms of coherence \cite{60-n, 61-n, 62-n, 63-n, 64-n}. In the analysis of the influence of the presence of the controllable noise level presented in preceding sub-sections, it is evident that some of the parameters quantifying performance of the BISON algorithm show better performance with the presence of noise.

As discussed (\S \ref{sec:evaluation:NodeVelocity}), overall movement is investigated using node--\textit{diffusion} \ensuremath{D} (Eq. \ref{eq:diffusion}) and --\textit{drift} \ensuremath{F} (Eq. \ref{eq:drift}),  the former representing the mean rate of change in average node velocity, the latter quantifying the evolution thereof.
The results for walled environments (bottom) and environments with
walls (top) are plotted in Fig. \ref{coefficient.combined}.
%

Similar plots 
from our other experiments all
show similar or stronger differences in the spread of the instantaneous values of the diffusion coefficient. We believe that this finding is important for the design of collective robot swarms (who will always operate under noise in the real world), and will be subject of further reports. Similar to the differences in the distribution of the diffusion coefficient values, there are notable quantitative and qualitative differences in the distribution of the drift coefficient, as shown above. By increasing the noise-level we cause the broadening of the distribution. A similar impact is visible between different environments (comparing (a) vs (c) and (b) vs (d) in either), indicating that drift and diffusion can be used to deduce some ranking of difficulty on our environments.

\subsection{Loss of Members and Collision-free Recovery}\label{subsec:results.loss.of.nodes}
\label{crossreference:exit.path.for.redundant.node}
\begin{figure}[h!]
    \centerline{\includegraphics[width=\hsize, height=0.95\hsize]{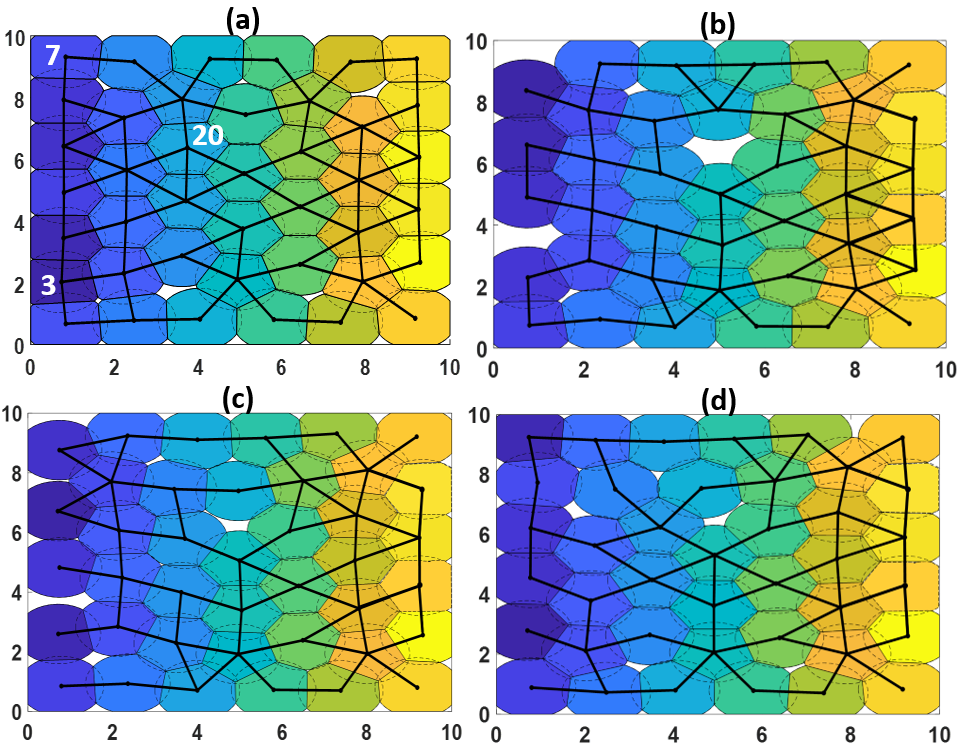}}
    \caption{(a) Snapshots of a WSN with the three nodes that are about to be lost unannounced labelled 3, 7, and 20. (b)-(d) visualizes the recovery process of the network to cover the holes generated by the loss.}
    \label{coverage-hole}
    \end{figure}

We also investigate the impact of node-loss as well as formulate a strategy to address it. As discussed in \S \ref{subsec:BISON.node.failure} we distinguish three types of nodes positions (side, corner and inside the collective), the nodes 3, 7 and 20, labelled in Fig. \ref{coverage-hole} are respective representatives thereof.

\begin{figure}[h!]
    \centerline{\includegraphics[width=\hsize]{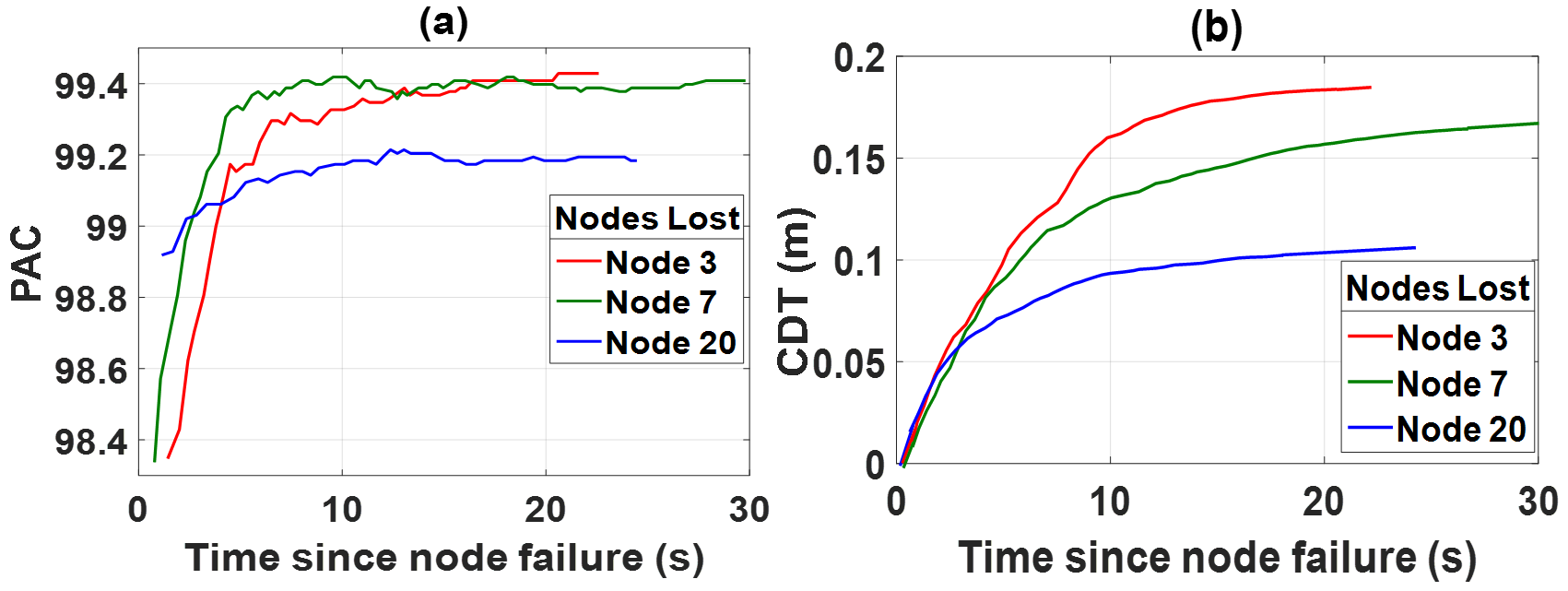}}
    \caption{The recovery of the network after node loss (cf Fig. \ref{coverage-hole}). Three different types of nodes (side, corner and center) were simulated to drop out (separately) and the resulting changes in coverage were recorded as well as the movement required to do so. Most of the lost coverage is recovered almost immediately and almost half the movement is incurred fine tuning the locations of the recovering nodes.}
    \label{exit}
    \end{figure}

Fig. \ref{coverage-hole} (a) shows a WSN suffering triple node loss (b) with BISON recovering from the resulting loss of coverage. We simulated individual node loss and recorded the evolution of both, \texttt{PAC} as well as \texttt{CDT} during the recovery of the remaining network \textit{after} the node-loss occurred. The results are plotted in Fig. \ref{exit}: we see a quick recovery of the network noting that the relative position within the network has a large impact on the recovery process and performance.

\begin{figure}[h!]
    \centerline{\includegraphics[width=0.6\hsize,height=0.6\hsize]{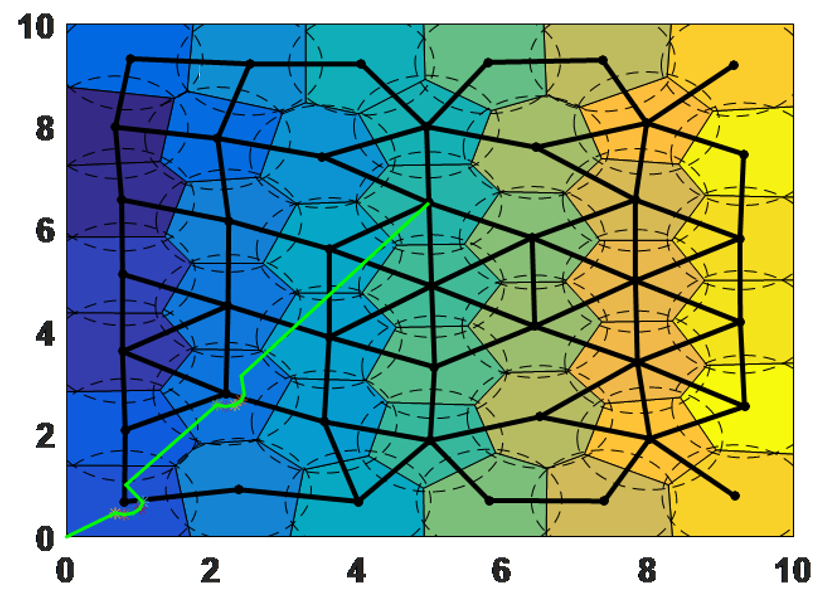}} 
    \caption{Simulated controlled node failure where a node pre-emptively removes itself from the network in case of e.g., low battery. The direct exit path taken is written in green, basic collision avoidance causes the node to circle around other nodes in its path.}
    \label{exit-path}
    \end{figure}

As part of our ongoing work, we start by simulating a short exit path for a randomly chosen node that lost most of its power, as shown in Fig. \ref{exit-path}.
Each node \textit{knows} the location of the starting point where nodes are injected from. A direct path is then planned from the current node location towards the injection point. During the motion of the node, it will sense if there is another node within its sensing range that blocks the path and will avoid it by moving around that node until it is back again on the planned path.

The green line in Fig. \ref{exit-path} illustrates the shortest exit path a certain low power sensor node follows to get out of the network before it dies out. We will further extend the work to consider the total time required to exit the area and the amount of energy consumed during the exit process, as well as include some approach to guide nodes along the shortest path in the network (in case that there are obstacles or blocking the shortest direct path).

\section{Conclusions and Future Work Plan}\label{sec:conclusion}
We presented BISON (Bio-inspired Self-organizing Network), a new Voronoi-based optimization algorithm for wireless sensor network deployment and reallocation in an unknown, obstacle-rich and noisy environment. One novel aspect of BISON is its ability to adapt to changes in the environment. A new method for node injection, guided by system requirements and based on satisfying coverage and connectivity conditions is proposed. Nodes calculate their Voronoi regions exclusively on the basis of locally available information and then continuously move towards the center of their region.  A number of performance criteria were introduced and used to validate the algorithm.\\
~\\
A summary of the algorithm results is the following:

\begin{enumerate}
\item In terms of the coverage and the number of nodes, BISON tends to achieve maximum coverage at an earlier time and with fewer nodes compared to the NSVA algorithm (with ratio of 1 to 3), under the same parameters. However, the average distance traveled is higher for BISON as new nodes enter the area of interest one at a time, while the NSVA algorithm tends to have a fixed number of nodes at the starting point waiting to be distributed in the network, implying that nodes need less movement during the deployment process.

\item The algorithm was also evaluated in a variety of environments that contain different obstacle shapes, scattering objects, and spacings, showing the broad applicability of BISON while performing well in a variety of case studies.

\item The presence of Gaussian noise allowed the sensor nodes to cover the region of interest faster than in a noise-free situation, and to reach the maximum coverage in shorter time.

\item The early nodes consume the most energy as they are the first to discover the region of interest and the optimum position is not yet clear to them. As more nodes enter the area, their movements become limited and the optimum position becomes better determined, leading to the decrease of energy consumption with time. Additionally, in a noisy environment, a trade-off between the coverage efficiency and the energy expenditure is defined.

\item Loss of some of the early nodes is analyzed, demonstrating robustness of the proposed BISON solution.

\item Lastly, the influence of noise is analyzed within drift-diffusion framework, lending an insight into future design of efficient deployment networks.
\end{enumerate}

Future work plan will consider the dynamics of the nodes' entry that could be somewhat improved, as the density near the entry point needs to be optimized. Variable sensing range for individual sensors will also add a layer of realistic implementation to the algorithm, as will the inclusion of temporally varied noise levels. Finally, a three-dimensional consideration of the algorithm need to be studied to better visualize the performance of the algorithm and have a better evaluation for real implementations.

\section*{Acknowledgments}
The authors acknowledge support from the UAE ICTFund grant ``Bio-inspired Self-organizing Services''.
AFI and DR acknowledge motivating discussion with Prof. V. Kumar (Univ. of Pennsylvania).


\section*{Biographies}
\subsubsection*{Khouloud Eledlebi} Biography will be added for final version.
\subsubsection*{Dymitr Ruta} Biography will be added for final version.

\subsubsection*{Hanno Hildmann}  Biography will be added for final version.
\subsubsection*{Fabrice Saffre}  Biography will be added for final version.
\subsubsection*{Yousof Al Hammadi}  Biography will be added for final version.
\subsubsection*{A.F. Isakovic}  Biography will be added for final version.
\section*{Supporting Online Materials}
Additional results, supporting the findings and statements in the text as well as some animations, are available from the corresponding author, upon reasonable request
\end{document}

%% file: diagram.tex
\begin{picture}(500,200)(-3, 0)
\linethickness{0.2mm}
\put(  0,200){
        \put(  0,   0){\line( 1, 0){180}}
        \put(  0, -80){\line( 1, 0){180}}
        \put(  0,   0){\line( 0,-1){ 80}}
        \put(180,   0){\line( 0,-1){ 80}}
    \put(  0,  0){
        \put(  3,-13) {\cite{15,16,21,36,37,42,46f} for}}
    \put(  5,-45){
        \put(  0, 25){
            \put( 8,-10) {Detecting holes}
            \put( 18,-20) {in coverage}}
        \put(  0,  25){\line( 0,-1){ 25}}
        \put( 80,  25){\line( 0,-1){ 25}}
        \put(  0,   0){\line( 1, 0){ 80}}
        \put( 80,  25){\line(-1, 0){ 80}}}
    \put( 95,-45){
        \put(  0, 25){
            \put(  3,-10) {Maintaining uni-}
            \put(  3,-20) {form distribution}}
        \put(  0,  25){\line( 0,-1){ 25}}
        \put( 80,  25){\line( 0,-1){ 25}}
        \put(  0,   0){\line( 1, 0){ 80}}
        \put( 80,  25){\line(-1, 0){ 80}}}
    \put( 43, -75){
        \put(  0, 25){
            \put( 13,-10) {Improving the }
            \put(  3,-20) {scheduling scheme}}
        \put(  0,  25){\line( 0,-1){ 25}}
        \put( 86,  25){\line( 0,-1){ 25}}
        \put(  0,   0){\line( 1, 0){ 86}}
        \put( 86,  25){\line(-1, 0){ 86}}}
        }
\put(  0,110){
        \put(  0,   0){\line( 1, 0){180}}
        \put(  0, -90){\line( 1, 0){180}}
        \put(  0,   0){\line( 0,-1){ 90}}
        \put(180,   0){\line( 0,-1){ 90}}
    \put(  0,  0){
        \put(  3,-13) {Leads to \cite{40,41,48,39,42a,43}}
        \put(  3,-23) {\cite{44,45,46,56-0,56,72}}}
    \put(  2,-30){
        \put(  17,-10) {\textbf{Maximising}}
        \put(  0,   0){\line( 0,-1){ 55}}
        \put( 86,   0){\line( 0,-1){ 55}}
        \put(  0,   0){\line( 1, 0){ 86}}
        \put( 86, -55){\line(-1, 0){ 86}}
        \put(  2,-15){
            \put(  17, -12) {\footnotesize{Area Coverage}}
            \put(  0,   0){\line( 0,-1){ 18}}
            \put( 82,   0){\line( 0,-1){ 18}}
            \put(  0,   0){\line( 1, 0){ 82}}
            \put( 82, -18){\line(-1, 0){ 82}}}
         \put(  2,-35){
            \put(  21, -12) {\footnotesize{Convergence}}
            \put(  0,   0){\line( 0,-1){ 18}}
            \put( 82,   0){\line( 0,-1){ 18}}
            \put(  0,   0){\line( 1, 0){ 82}}
            \put( 82, -18){\line(-1, 0){ 82}}}}
    \put( 92,-30){
        \put(  17,-10) {\textbf{Minimizing}}
        \put(  0,   0){\line( 0,-1){ 55}}
        \put( 86,   0){\line( 0,-1){ 55}}
        \put(  0,   0){\line( 1, 0){ 86}}
        \put( 86, -55){\line(-1, 0){ 86}}
        \put(  2,-15){
            \put(  9, -12) {\footnotesize{Energy consumption}}
            \put(  0,   0){\line( 0,-1){ 18}}
            \put( 82,   0){\line( 0,-1){ 18}}
            \put(  0,   0){\line( 1, 0){ 82}}
            \put( 82, -18){\line(-1, 0){ 82}}}
         \put(  2,-35){
            \put(  19, -8) {\footnotesize{Computational}}
            \put(  25,-15) {\footnotesize{complexity}}
            \put(  0,   0){\line( 0,-1){ 18}}
            \put( 82,   0){\line( 0,-1){ 18}}
            \put(  0,   0){\line( 1, 0){ 82}}
            \put( 82, -18){\line(-1, 0){ 82}}}}
        }
\put(200,170){
\linethickness{0.5mm}
    \put(  0, 30){
        \put(  6,-13) {Voronoi diagrams in}
        \put(  12,-23) {WSN deployment}}
    \put(  0,  0){
        \put(  0,  0){
            \put(0,0){\line( 1, 0){100}}
            \put(0,0){\line( 0, 1){ 30}}}
        \put(100,  0){
            \put(0,0){\line(-1, 0){100}}
            \put(0,0){\line( 0, 1){ 30}}}
        \put(  0, 30){
            \put(0,0){\line( 1, 0){100}}
            \put(0,0){\line( 0,-1){ 30}}}
        \put(100, 30){
            \put(0,0){\line(-1, 0){100}}
            \put(0,0){\line( 0,-1){ 30}}}}}
\linethickness{0.2mm}
\put(190,150){
        \put(  0,   0){\line( 1, 0){120}}
        \put(  0,-143){\line( 1, 0){120}}
        \put(  0,   0){\line( 0,-1){143}}
        \put(120,   0){\line( 0,-1){143}}
    \put(  0,  0){
        \put(  3,-13) {Generated through \cite{16}}
        \put(  3,-23) {\cite{40, 41,38,39,44}}
        \put(  3,-33) {\cite{46f,46ba,46bb}}}
        \put(  2,-40){
        \put(  3,-10) {Intersection of neighbours'}
        \put( 11,-20) {perpendicular bisectors}
        \put(  0,   0){\line( 0,-1){ 70}}
        \put(116,   0){\line( 0,-1){ 70}}
        \put(  0,   0){\line( 1, 0){116}}
        \put(116, -70){\line(-1, 0){116}}
        \put(  2,-25){
            \put( 25, -8) {\footnotesize{Euclidean distance}}
            \put( 20,-15) {\footnotesize{(homogeneous nodes)}}
            \put(  0,   0){\line( 0,-1){ 18}}
            \put(112,   0){\line( 0,-1){ 18}}
            \put(  0,   0){\line( 1, 0){112}}
            \put(112, -18){\line(-1, 0){112}}}
         \put(  2,-45){
            \put( 26, -8) {\footnotesize{Laguerre distance}}
            \put( 20,-15) {\footnotesize{(heterogeneous node)}}
            \put(  0,   0){\line( 0,-1){ 18}}
            \put(112,   0){\line( 0,-1){ 18}}
            \put(  0,   0){\line( 1, 0){112}}
            \put(112, -18){\line(-1, 0){112}}}}
        \put(  5,-120){
            \put( 20, -8) {\footnotesize{Intersected with node's}}
            \put( 35,-15) {\footnotesize{sensing range}}
            \put(  0,   0){\line( 0,-1){ 18}}
            \put(112,   0){\line( 0,-1){ 18}}
            \put(  0,   0){\line( 1, 0){112}}
            \put(112, -18){\line(-1, 0){112}}}
}
\put(320,200){
        \put(  0,   0){\line( 1, 0){180}}
        \put(  0,-110){\line( 1, 0){180}}
        \put(  0,   0){\line( 0,-1){110}}
        \put(180,   0){\line( 0,-1){110}}
    \put(  0,  0){
        \put(  3,-13) {Implemented on randomly distributed}
        \put(  3,-23) {homogeneous / heterogeneous nodes in}
        \put(  3,-33) {known / unknown environments in \cite{40}}
        \put(  2,-44) {\cite{41,42,44,45,46,46f,46ba,46bb}}}
    \put(  2,-50){
        \put(  10,-10) {\textbf{Avoiding noise}}
        \put(  0,   0){\line( 0,-1){ 55}}
        \put( 86,   0){\line( 0,-1){ 55}}
        \put(  0,   0){\line( 1, 0){ 86}}
        \put( 86, -55){\line(-1, 0){ 86}}
        \put(  2,-15){
            \put(  22, -8) {\footnotesize{Addition of}}
            \put(  14,-15) {\footnotesize{random numbers}}
            \put(  0,   0){\line( 0,-1){ 18}}
            \put( 82,   0){\line( 0,-1){ 18}}
            \put(  0,   0){\line( 1, 0){ 82}}
            \put( 82, -18){\line(-1, 0){ 82}}}
         \put(  2,-35){
            \put(  25, -8) {\footnotesize{Pre-defined}}
            \put(  21,-15) {\footnotesize{anchor nodes}}
            \put(  0,   0){\line( 0,-1){ 18}}
            \put( 82,   0){\line( 0,-1){ 18}}
            \put(  0,   0){\line( 1, 0){ 82}}
            \put( 82, -18){\line(-1, 0){ 82}}}}
    \put( 92,-50){
        \put(  3,-10) {\textbf{Avoiding obstacles}}
        \put(  0,   0){\line( 0,-1){ 55}}
        \put( 86,   0){\line( 0,-1){ 55}}
        \put(  0,   0){\line( 1, 0){ 86}}
        \put( 86, -55){\line(-1, 0){ 86}}
        \put(  2,-15){
            \put(  15, -12) {\footnotesize{Repulsive force}}
            \put(  0,   0){\line( 0,-1){ 18}}
            \put( 82,   0){\line( 0,-1){ 18}}
            \put(  0,   0){\line( 1, 0){ 82}}
            \put( 82, -18){\line(-1, 0){ 82}}}
         \put(  2,-35){
            \put(  20, -8) {\footnotesize{Scanning the}}
            \put(  21,-15) {\footnotesize{entire region}}
            \put(  0,   0){\line( 0,-1){ 18}}
            \put( 82,   0){\line( 0,-1){ 18}}
            \put(  0,   0){\line( 1, 0){ 82}}
            \put( 82, -18){\line(-1, 0){ 82}}}}
        }
\put(320, 80){
        \put(  0,   0){\line( 1, 0){180}}
        \put(  0, -80){\line( 1, 0){180}}
        \put(  0,   0){\line( 0,-1){ 80}}
        \put(180,   0){\line( 0,-1){ 80}}
    \put(  0,  0){
        \put(  3,-13) {Position \cite{16,40,38,39,44,46bb}}}
    \put(  5,-45){
        \put(  0, 25){
            \put( 38,-10) {\textbf{Voronoi ploygonial}}
            \put( 28,-20) {edges, vertices, centroids}}
        \put(  0,  25){\line( 0,-1){ 25}}
        \put(170,  25){\line( 0,-1){ 25}}
        \put(  0,   0){\line( 1, 0){170}}
        \put(170,  25){\line(-1, 0){170}}}
    \put(  5,-75){
        \put(  0, 25){
            \put( 5,-10) {\textbf{Voronoi edges'} midpoint intersection /}
            \put( 18,-20) {closest to the furthest point vertex}}
        \put(  0,  25){\line( 0,-1){ 25}}
        \put(170,  25){\line( 0,-1){ 25}}
        \put(  0,   0){\line( 1, 0){170}}
        \put(170,  25){\line(-1, 0){170}}}
        }
\put(185,180){\ensuremath{\Leftarrow}}
\put(246,157){\ensuremath{\Downarrow}}
\put(305,180){\ensuremath{\Rightarrow}}
\put( 86,113){\ensuremath{\Downarrow}}
\put(246, 33){\ensuremath{\Downarrow}}
\put(405,83){\ensuremath{\Downarrow}}
\end{picture}

%% file: arXiv.DRAFT.September.2020.bbl
\begin{thebibliography}{1}
\bibitem{4}
P. Tiwari, V. P. Saxena, R. G. Mishra and D. Bhavsar, ``Wireless Sensor Networks: Introduction, Advantages, Applications and Research Challenges'', Open International Journal of Technology Innovations and Research, vol. 14, pp. 1-11, 2015.

\bibitem{6}
L. Poudyal and B. Sen, ``A Survey on Localization and Covering Techniques in Wireless Sensor Networks,'' International Journal of Computer Applications, vol. 67, no. 7, pp. 23-27, 2013.

\bibitem{9}
D. Ye, D. Gong and W. Wang, ``Application of Wireless Sensor Networks in Environmental Monitoring,'' in 2nd IEEE International Conference on Power Electronics and Intelligent Transportation System, Shenzhen, China, December 2009.

\bibitem{1}
F. Guerriero, A. Violi, E. Natalizio, V. Loscri and C. Costanzo, ``Modelling and Solving Optimal placement Problems in Wireless Sensor Networks,'' Applied Mathematical Modelling, vol. 35, no. 1, pp. 230-241, 2011.

\bibitem{2}
D. Chandirasekaran and T. Jayabarathi, ``A Case Study of Bio-Optimization Techniques for Wireless Sensor Network in Node Location Awareness,'' Indian Journal of Science and Technology, vol. 8, no. 31, pp. 1-9, 2015.

\bibitem{3}
H. Ghayvat, S. Mukhopadhyay, X. Gui and N. Suryadevara, ``WSN- and IoT-Based Smart Homes and Their Extension to Smart Buildings,'' Sensors, vol. 15, pp. 10350-10379, 2015.

\bibitem{10}
Z. Chuan, Z. Chunlin, S. Lei and H. Guangjie, ``A Survey on Coverage and Connectivity Issues in Wireless Sensor Networks,'' Journal of Network and Computer Applications, vol. 35, no. 2, pp. 619-632, 2012.

\bibitem{11}
P. Sahu, and S. R. Gupta, ``Deployment Techniques in Wireless Sensor Networks,'' International Journal of Soft Computing and Engineering, vol. 2, no. 3, pp. 525-526, 2012.

\bibitem{14}
S. Mnasri, N. Nasri and T. Val, ``The Deployment in the Wireless Sensor Networks: Methodologies, Recent Works and Applications,'' in International Conference on Performance Evaluation and Modeling in Wired and Wireless Networks (PEMWN 2014), Nov 2014, Sousse, Tunisia, Proceedings of PEMWN 2014 .

\bibitem{13}
M. Kiran, K. Kamal, G. Nitin, ``Application-based Study on Wireless Sensor Network,'' International Journal of Computer Applications, vol. 21, no. 8, pp. 9-15, 2011.

\bibitem{15}
M. Abbasi, M. S. Abd Latif and H. Chizari, ``Bioinspired Evolutionary Algorithm Based for Improving Network Coverage in Wireless Sensor Networks,'' The Scientific World Journal, vol. 2014, pp. 1-8, 2014.

\bibitem{16}
Q. Du, V. Faber and M. Gunzburger, ``Centroidal Voronoi Tessellations: Applications and Algorithms,'' SIAM Review, vol. 41, no. 4, pp. 637-676, 1999.

\bibitem{17}
K. Eledlebi, D. Ruta, F. Saffre, Y. Al Hammadi and A. F. Isakovic, ``A Model for Self-deployment of Autonomous Mobile Sensor Network in an Unknown Indoor Environment,'' in Ad Hoc Networks: 9th International Conference (AdHocNets 2017), Proceedings, Niagara Falls, ON, Canada, Springer, pp. 208-215, 2017.

\bibitem{21}
M. A. Adnan, M. Abdur Razzaque, I. Ahmed and I. F. Isnin, ``Bio-mimic Optimization Strategies in Wireless Sensor Networks: A Survey,'' Sensors, vol. 14, pp. 299-345, 2014.

\bibitem{22}
T. Sheltami, A. Mahmoud, K. Alsafari and E. shakshuki, ``Self-Organizing Sensor Networks Coverage Problem,'' in IEEE 26th Biennial Symposium on Communications (QBSC), Kingston, ON, Canada, May 2012.

\bibitem{23}
T. Watteyne, ``Energy-efficient Self-organization in Wireless Sensor Networks,'' PhD thesis manuscript, INSA Lyon, 2008.

\bibitem{24}
A. Norouzi and A. H. Zaim, ``Genetic Algorithm Application in Optimization of Wireless Sensor Networks,'' The Scientific World Journal, vol. 2014, pp. 1-15, 2014.

\bibitem{36}
R. Klein, ``Voronoi Diagrams and Delaunay Triangulations,'' in Encyclopedia of Algorithms, NewYork, Springer Science and Business Media, ISBN: 9789814447645, pp. 1-5, 2014.

\bibitem{37}
M. Hasegawa and T. Masaharu, ``On the Pattern of Space Division by Territories,'' Annuals of the Institute of Statistical Mathematics, vol. 28, no. 1, pp. 509-519, December 1976.

\bibitem{40}
H.-J. Lee, Y.-H. Kim, Y.-H. Han and C. Y. Park, ``Centroid-based Movement Assisted Sensor Deployment Schemes in Wireless Sensor Networks,'' in Vehicular Technology Conference, USA, Fall 2010.

\bibitem{41}
A. Pietrabissa, F. Liberati and G. Oddi, ``A Distributed Algorithm for Ad-hoc Network Partitioning Based on Voronoi Tessellation,'' Elsevier: Ad Hoc Networks, vol. 46, pp. 37-47, 2016.

\bibitem{42}
M. A. M. Vieira, L. F. M. Vieira, L. B. Ruiz, A. Loureiro, A. Fernandes and J. Nougueira, ``Scheduling Nodes in Wireless Sensor Networks: A Voronoi Approach,'' in 28th Annual IEEE International Conference on Local Computer Networks, 2003. LCN '03. Proceedings., Bonn/Konigswinter, Germany, 2003.

\bibitem{28}
O. Banimelhem, M. Mowafi and W. Aljoby, ``Genetic Algorithm Based Node Deployment in Hybrid Wireless Sensor Network,'' Communications and Network, vol. 5, pp. 273-279, 2013.

\bibitem{29}
S. Kaur and R. Uppal, ``Dynamic Deployment of Homogeneous Sensor Nodes Using Genetic Algorithm with Maximum Coverage,'' in 2nd International Conference on Computing for Sustainable Global Development, New Delhi, India, 2015.

\bibitem{32}
Yipeng Qu and Stavros V. Georgakopoulos, ``A Centralized Algorithm for Prolonging the Lifetime of Wireless Sensor Networks using Particle Swarm Optimization,'' in IEEE Wireless and Microwave Technology Conference (WAMICON), Florida, USA, April 2012.

\bibitem{48}
N. Rahmani, F. Nematy, A. M. Rahmani and M. Hosseinzadeh, ``Node Placement for Maximum Coverage Based on Voronoi Diagram Using Genetic Algorithm in Wireless Sensor Networks,'' Australian Journal of Basic and Applied Sciences, vol. 5, no. 12, pp. 3221-3232, 2011.

\bibitem{51}
N. A. B. Ab Aziz, A. W. Mohemmed and B. D. Sagar, ``Particle Swarm Optimization and Voronoi Diagram for Wireless Sensor Networks Coverage Optimization,'' in IEEE International Conference on Intelligent and Advanced Systems, Kuala Lumpur, Malaysia, November 2007.

\bibitem{38}
G. Wang, G. Cao and T. L. Porta, ``Movement-Assisted Sensor Deployment,'' IEEE Transactions on Mobile Computing, vol. 5, no. 6, pp. 640-652, 2006.

\bibitem{IEEE-j1}
D. Wang, B. Xie and D. P. Agrawali, ``Coverage and Lifetime Optimization Of Wireless Sensor Networks With Gaussian Distribution,'' IEEE Transactions on mobile computing, vol. 7, no. 12, pp. 1444-1458,  2008.

\bibitem{39}
M. R. Senouci, A. Mellouk, K. Asnoune and F. Y. Bouhidel, ``Movement-Assisted Sensor Deployment Algorithms: A Survey and Taxonomy,'' IEEE Communication Surveys and Tutorials, vol. 17, no. 4, pp. 2493-2510, 2015.

\bibitem{42a}
J. Stergiopoulos and A. Tzes, ``Voronoi-based Coverage Optimization for Mobile Networks with Limited Sensing Range - A Directional Search Approach,'' in American Control Conference, Missouri, USA, 2009.

\bibitem{43}
J. Zou, J. Kusyk, M. U. Uyar, S. Gundry and C. S. Sahin, ``Bio-inspired and Voronoi-based Algorithms for Self-positioning Autonomous Mobile Nodes,'' in IEEE Military Communications Conference (MILCOM 2012), Orlando, FL, USA, October 2009.

\bibitem{44}
J. Zou, S. Gundry, J. Kusyk, C. S. Sahin and M. U. Uyar, ``Bio-inspired and voronoi-based algorithms for self-positioning of autonomous vehicles in noisy environments,'' in Proceedings of the 8th International Conference on Bio-inspired Information and Communications Technologies, Boston, Massachusetts, pp. 12-22, Dec 2014.

\bibitem{45}
J. Kusyk, J. Zou, S. Gundry, C. S. Sahin and M. U. Uyar, ``Metrics for Performance Evaluation of Self-positioning Autonomous MANET Nodes,'' in 35th IEEE Sarnoff Symposium, Newark, NJ, USA, June 2012.

\bibitem{46}
S. Gundry, J. Zou, C. S. Sahin, J. Kusyk and M. U. Uyar, ``Autonomous and Fault Tolerant Vehicular Self Deployment Mechanisms in MANETs,'' in IEEE International Conference on Technologies for Homeland Security, Waltham, MA, USA, November 2013.

\bibitem{46a}
N. Bartolini, T. Calamoneri, T. F. La Porta and S. Silvestri, ``Autonomous Deployment of Heterogeneous Mobile Sensors,'' IEEE Transactions on Mobile Computing, vol. 10, no. 6, pp. 753-766, 2011.

\bibitem{46b}
Y. Kantaros, M. Thanou and A. Tzes, ``Distributed Coverage Control for Concave Areas by a Heterogeneous Robot-swarm with Visibility Sensing Constraints,'' Elsevier: Automatica, vol. 53, pp. 195-207, 2015.

\bibitem{46c}
H. Imai, M. Iri and K. Murota, ``Voronoi Diagram in the Laguerre Geometry and its Applications,'' Society for Industrial and Applied Mathematics, vol. 14, no. 1, pp. 93-105, 1985.

\bibitem{46cc}
H. Mahboubi and A. G. Aghdam, ``Distributed Deployment Algorithms for Coverage Improvement in a Network of Wireless Mobile Sensors: Relocation by Virtual Force,'' IEEE Transactions on Control of Network Systems, vol. PP, no. 99, pp. 1-19, 2016.

\bibitem{46d}
H. Mahboubi, K. Moezzi, A. G. Aghdam and K. Sayrafian-Pour, ``Distributed Sensor Coordination Algorithms for Efficient Coverage in a Network of Heterogeneous Mobile Sensors,'' IEEE Transactions on Automatic Control, vol. 62, no. 11, pp. 5954-5961, 2017.

\bibitem{46f}
N. Bartolini, S. Ciavarella, S. Silvestri and T. La Porta, ``On the Vulnerabilities of Voronoi-based Approaches to Mobile Sensor Deployment,'' IEEE Transactions on Mobile Computing, vol. 15, no. 12, pp. 3114-3128, 2016.

\bibitem{46ba}
N. Bartolini, T. Calamoneri, E. G. Fusco, A. Massini and S. Silvestri, ``Autonomous Deployment of Self-organizing Mobile Sensors for a Complete Coverage,'' Springer: International Workshop on Self-organizing Systems, pp. 194-205, 2008.

\bibitem{46bb}
N. Bartolini, T. Calamoneri, S. Ciavarella, T. La Porta and S. Silvestri, ``Autonomous Mobile Sensor Placement in Complex Environments,'' ACM Transactions on Autonomous and Adaptive Systems, vol. 12, no. 2, pp. 1-28, 2017.

\bibitem{58}
L. Liu, G. Han, H. Wang and J. Wan, ``Obstacle-avoidance Minimal Exposure Path for Heterogeneous Wireless Sensor Networks,'' Ad Hoc Networks Journal, vol. 55, pp. 50-61, 2016

\bibitem{cortes}
J. Cortes, S. Martinez, T. Karatas and F. Bullo, ``Coverage Control for Mobile Sensing Networks,'' IEEE Transactions on Robotics and Automation, vol. 20, no. 2, pp. 243-255, 2004.

\bibitem{57}
S. L. Mohammed, ``Distance Estimation Based on RSSI and Log-normal Shadowing Models for ZigBee Wireless Sensor Network,'' Engineering and Technology Journal, vol. 34, no. 15, pp. 2950-2960, 2016.

\bibitem{IEEE-j2}
H. Chenji and R. Stoleru, ``Toward Accurate Mobile Sensor Network Localization in Noisy Environments,'' IEEE Transactions on mobile computing, vol. 12, no. 6, pp. 1-14, June 2013.

\bibitem{56-0}
H. Zhang and J. C. Hou, ``Maintaining Sensing Coverage and Connectivity in Large Sensor Networks,'' Ad Hoc and Sensor Wireless Networks Journal, vol. 1, no. 1--2, pp. 89--124, 2005.

\bibitem{56}
Y.-C. Wang, C.-C. Hu and Y.-C. Tseng, ``Efficient Deployment Algorithms for Ensuring Coverage and Connectivity of Wireless Sensor Networks,'' in First International Conference on Wireless Internet,  Budapest, Hungary, July 2005.

\bibitem{46g}
A. Ohya, T. Ohno and S. Yuta, ``Obstacle Detectability of Ultrasonic Ranging System and Sonar Map Understanding,'' Elsevier: Robotics and Autonomous Systems, vol. 18, no. 1-2, pp. 251-257, 1996.

\bibitem{60-n}
C. A. Yates, R. Erban, C. Escudero, I. D. Couzin, J. Buhl, I. G. Kevrekidis, P. K. Maini and D. J. T. Sumpter, ``Inherent Noise Can Facilitate Coherence in Collective Swarm Motion,'' Proceedings of the National Academy of Sciences in the United States of America, vol. 106, no. 14, pp. 5464-5469, 2009.

\bibitem{62}
G. Nimisha, B. Indrajit and S. Tuhina, ``Energy Efficient Coverage of Static Sensor Nodes Deciding on Mobile Sink Movements Using Game Theory,'' in IEEE Conference on Applications and Innovations in Mobile Computing, Kolkata, February 2014.

\bibitem{72}
M. Abo-Zahhad, N. Sabor, S. Sasaki and S. M. Ahmed, ``A Centralized Immune-Voronoi Deployment Algorithm for Coverage Maximization and Energy Conservation in Mobile Wireless Sensor Networks,'' Information Fusion Journal, vol. 30, pp. 36-51, 2015.

\bibitem{55}
A. Khelil and R. Beghdad, ``ESA: an Efficient Self-deployment Algorithm for Coverage,'' in The 7th International Conference on Emerging Ubiquitous Systems and Pervasive Networks, London, UK, September 2016.

\bibitem{61-n}
J. Parrish, S. V. Viscido and D. Grunbaum, ``Self-Organized Fish Schools: An Examination of Emergent Properties,'' The Biological Bulletin Journal, vol. 202, no. 3, pp. 296-305, 2002.

\bibitem{62-n}
V. Gabor, C. Viragh, G. Somorjai, N. Tarcai, T. Szorenyi, T. Nepusz and T. Vicsek, ``Outdoor Flocking and Formation Flight with Autonomous Aerial Robots,'' in IEEE International Conference on Intelligent Robots and Systems, 2014.

\bibitem{63-n}
N. W. F. Bode, D. W. Franks and A. J. Wood, ``Making Noise: Emergent Stochasticity in Collective Motion,'' Journal of Theoretical Biology, vol. 267, no. 3, pp. 292-299, 2010.

\bibitem{64-n}
B. Blonder, T. A. Wey, A. Dornhaus, R. James and A. Sih, ``Temporal Dynamics and Network Analysis,'' Journal of Methods in Ecology and Evolution, vol. 3, no. 6, pp. 958-972, 2012.

\bibitem{vbook}
F. Aurenhammer, R. Klein and D.-T. Lee, ``Voronoi Diagrams and Delaunay Triangulations,'' Singapore: World Scientific, 2013.

\end{thebibliography}
